\documentclass[twocoloumn]{aa} 

\usepackage{graphicx}
\usepackage{txfonts}

\usepackage{xcolor}
\usepackage[colorlinks]{hyperref}
\hypersetup{ colorlinks, linkcolor=blue, citecolor=blue }
\usepackage{ulem}
\usepackage{nicefrac}

\definecolor{darkgreen}{rgb}{0,0.6,0}

\begin{document} 

\title{Observational signatures of the surviving donor star in the double detonation model of Type Ia supernovae}

\titlerunning{Surviving donors of DDet SNe Ia}
\authorrunning{Z.-W. Liu, et al.}

\author{Zheng-Wei Liu\inst{1,2,3}
\and
Friedrich K. R\"{o}pke\inst{4,5}
\and 
Yaotian Zeng\inst{1,2,3} 
\and
Alexander Heger\inst{6,7,8,9}
}

\institute{
Yunnan Observatories, Chinese Academy of Sciences, 396 Yangfangwang, Guandu District, Kunming 650216, P.R. China\\
              \email{zwliu@ynao.ac.cn}
\and
Key Laboratory for the Structure and Evolution of Celestial Objects, CAS, Kunming 650216, P.R. China
\and
University of Chinese Academy of Science, Beijing 100012, P.R. China
\and 
Zentrum f{\"u}r Astronomie der Universit{\"a}t Heidelberg, Institut f{\"u}r Theoretische Astrophysik, Philosophenweg 12, 69120 Heidelberg, Germany
\and
Heidelberger Institut f{\"u}r Theoretische Studien, Schloss-Wolfsbrunnenweg 35, 69118 Heidelberg, Germany
\and
School of Physics and Astronomy, Monash University, 19 Rainforest Walk, VIC 3800, Australia
\and
Australian Research Council Centre of Excellence for Gravitational Wave Discovery (OzGrav), Clayton, VIC 3800, Australia
\and
Center of Excellence for Astrophysics in Three Dimensions (ASTRO-3D), Stromlo, ACT 2611, Australia
\and
Joint Institute for Nuclear Astrophysics, National Superconducting Cyclotron Laboratory, Michigan State University, 1 Cyclotron Laboratory, East Lansing, MI 48824-1321, USA
}

\abstract{
The sub-Chandrasekhar mass double-detonation (DDet) scenario is a  contemporary model for Type Ia supernovae (SNe Ia).  The donor star in the DDet scenario is expected to survive the explosion and to be ejected at the high orbital velocity of a compact binary system. For the first time, we consistently perform three-dimensional (3D) hydrodynamical simulations of the interaction of SN ejecta with a helium (He) star companion within the DDet scenario.  We map the outcomes of 3D impact simulations into one-dimensional stellar evolution codes and follow the long-term evolution of the surviving He-star companions. Our main goal is to provide the post-impact observable signatures of surviving He-star companions of DDet SNe Ia, which will support the search for such companions in future observations.  Such surviving companions are ejected with high velocities of up to about $930\,\rm{km\,s^{-1}}$.  We find that our surviving He-star companions become significantly overluminous for about $10^{6}\,\mathrm{yr}$ during the thermal re-equilibration phase. After the star re-establishes thermal equilibrium, its observational properties are not sensitive to the details of the ejecta-donor interaction.  We apply our results to hypervelocity star US~708, which is the fastest unbound star in our Galaxy, travelling with a velocity of about 1,200$\,\mathrm{km\,s^{-1}}$, making it natural candidate for an ejected donor remnant of a DDet SN Ia. We find that a He-star donor with an initial mass of $\gtrsim0.5\,\mathrm{M}_{\odot}$ is needed to explain the observed properties of US~708.  Based on our detailed binary evolution calculations, however, the progenitor system with such a massive He-star donor cannot get close enough at the moment of SN explosion to explain the high velocity of US~708.  Instead, if US~708 is indeed the surviving He-star donor of a DDet SN~Ia, it would require the entire pre-SN progenitor binary to travel at a velocity of about $400\,\mathrm{km\,s^{-1}}$.  It could, for example, have been ejected from a globular cluster in the direction of the current motion of the surviving donor star.
} 


\maketitle
%

\section{INTRODUCTION}
\label{sec:introduction}

It is generally accepted that Type Ia supernovae (SNe~Ia) result from thermonuclear explosions of white dwarfs (WDs) in interacting binary systems.  SNe Ia have been used as accurate cosmic distance indicators, which led to the discovery of the accelerating expansion of the Universe \citep{Ries98, Schmidt98, Perl99}.  The specific progenitor systems of SNe Ia and their explosion mechanism, however, remain an unsolved problem.  Different progenitor scenarios have been proposed for explaining SNe~Ia \citep[][]{Hillebrandt2013, Maoz2014}.

The sub-Chandrasekhar mass double-detonation (DDet) scenario is a promising model to explain normal SNe Ia \citep[e.g.,][]{Shen2018, Townsley2019, Gronow2020, Gronow2021}.  In this scenario, a WD accretes material from its He-rich companion star, which could be a He-burning star or a He WD, to accumulate a He layer on its surface.  If the He shell reaches a critical mass, $\simeq0.02$--$0.2\,\rm{M_{\odot}}$ \citep{Woosley2011, Neunteufel2016,Polin2019}, it triggers an initial detonation that ignites a second detonation of the core material.  The entire sub-Chandrasekhar mass WD then undergoes a thermonuclear explosion \citep[e.g.,][]{Taam1980, Woosley1986,Livn95, Fink07, Fink10, Sim10, Moll13, Gronow2020, Gronow2021, Boos2021}.  The DDet scenario has recently gained attention because of its attractive features in explaining current observations for SNe Ia.  For instance, the lack of H emission in SN Ia spectra \citep{Leon07} and the non-detection of a pre-explosion companion in HST imaging \citep{Li11} are inherent to DDet models.  Early UV flash signatures of ejecta interaction with a companion \citep{Kase10} should be very small in the DDet scenario because binary systems are generally very close at the moment of SN explosions.  In addition, population synthesis calculations show that the DDet scenario can explain a large fraction of SNe Ia as well as their delay time distribution (DTD, \citealt{Ruit11, Ruiter2014}).

\begin{figure*}
  \begin{center}
    {\includegraphics[width=0.48\textwidth, angle=0]{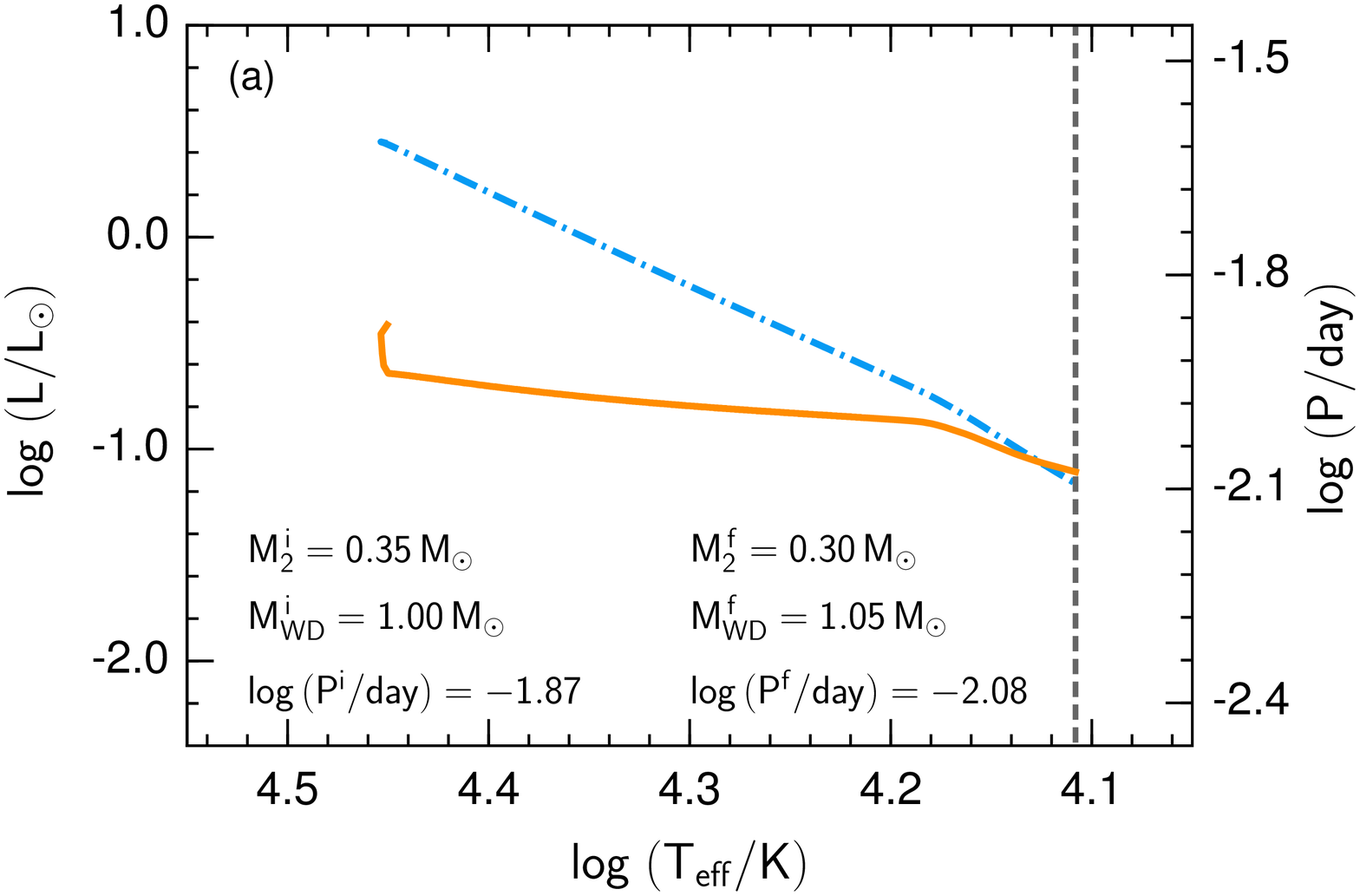}}
    \hfill
    {\includegraphics[width=0.48\textwidth, angle=0]{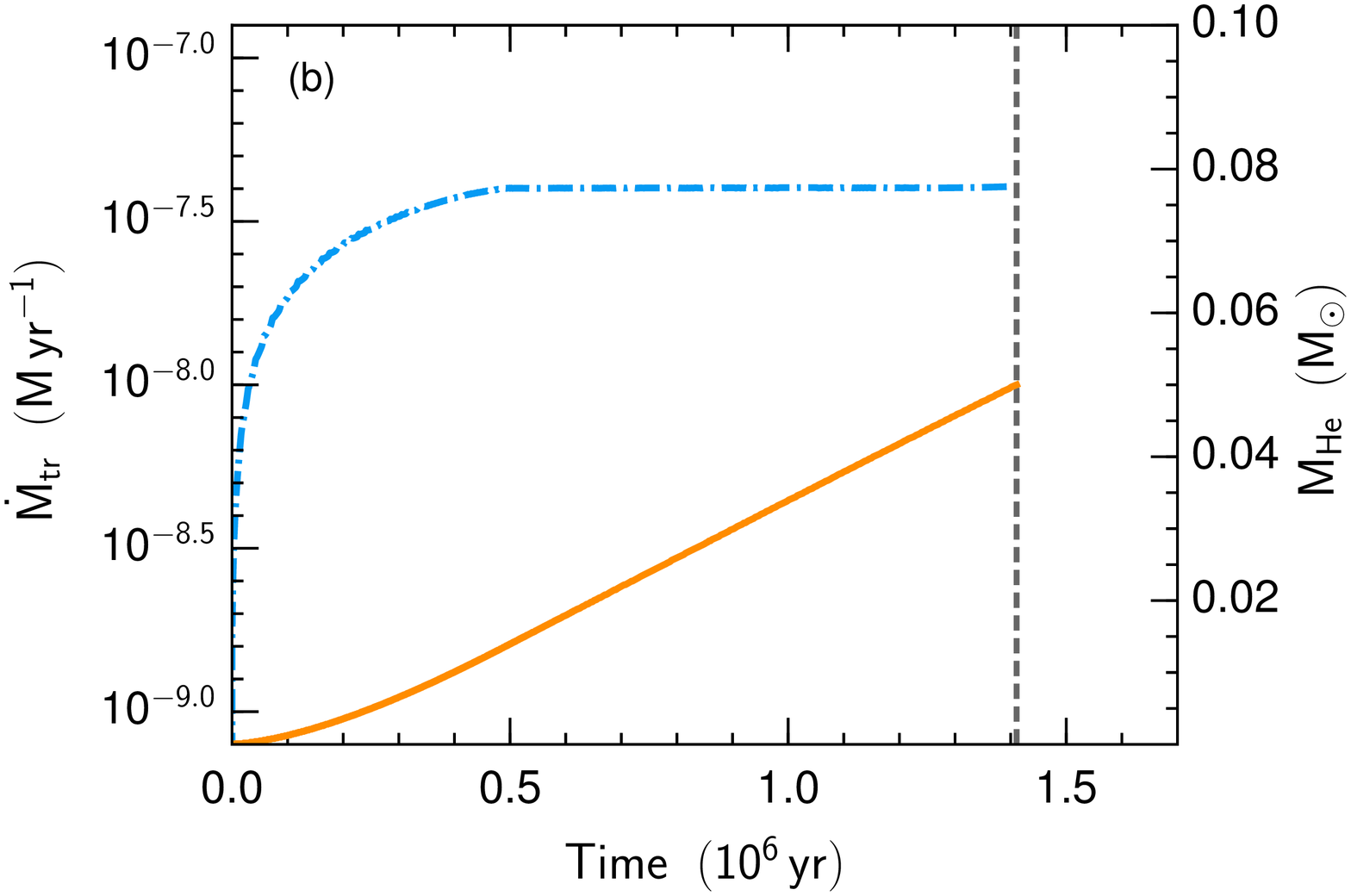}}

    \caption{\textbf{Left Panel:} evolutionary tracks of the luminosity of He-star companion (\textsl{cyan dash-dotted line}) and the orbital period of the binary system (\textsl{orange solid line}) during the whole mass-transfer process until the SN Ia explosion.  \textbf{Right Panel:} time evolution of the mass transfer rate ($\dot M_{\mathrm{tr}}$, \textsl{cyan dash-dotted line}) and the mass-growth of a helium shell onto the WD ($M_{\mathrm{He}}$, \textsl{orange solid line}). The vertical dotted lines give the moment of the SN explosion. The initial and final binary parameters (WD mass $M_{\mathrm{WD}}$, companion mass $M_{\mathrm{2}}$, orbital period $P$) in our detailed binary calculations are shown in the plot and indicated with  superscript letters `$\mathrm{i}$' and `$\mathrm{f}$', respectively.}

\label{Fig:star1}
  \end{center}
\end{figure*}

Observations of several binary systems composed of a WD and a He-rich companion star, e.g., KPD~1930+2752, V445~Pup, HD~49798 and CD-$30^{\circ}11223$ systems \citep{Maxt00, Geie07, Kato03, Kudr78, Venn12, Geie13} seem to support this scenario.  For instance, CD-$30^{\circ}11223$ is a WD~+~sdB star system with a WD mass of $M_{\rm{WD}}=0.76\,\rm{M_{\odot}}$, a companion mass of $M_{\rm{sdB}}=0.51\,\rm{M_{\odot}}$, and a short orbital period of $P_{\!\!\mathrm{orb}}\simeq1.2$ hours.  \citet{Venn12} and \citet{Geie13} suggest that CD-$30^{\circ}11223$ will likely explode as a SN Ia via the DDet scenario during its future evolution.

In the DDet scenario, the companion star is expected to be significantly shocked and heated while parts of its outer layers are removed during the ejecta-companion interaction. The binary system is disrupted, but the companion star survives the interaction with the supernova blast wave and is ejected at high speed that is dominated by its pre-explosion orbital velocity \citep[e.g.,][]{Whee75, Fryxell1981, Livne1992,Mari00, Pakm08, Liu2012, Liu2013a,Liu2013b,Liu2013c,Pan12,Bauer2019,Zeng2020}. Binary progenitor systems in the DDet scenario are relatively close when SN Ia explosions go off.  In this scenario, the companion stars are expected to have a high orbital velocity of up to $\simeq900\,\rm{km\,s^{-1}}$ at the moment of explosion \citep[e.g.,][]{Geie15, Neunteufel2020}. The surviving companion stars from this scenario are therefore good candidates of hypervelocity stars such as US~708 \citep[e.g.,][]{Geie13, Geie15, Neunteufel2020,Neunteufel2021}.  An unambiguous identification of a donor remnant of such an event would therefore support the DDet scenario of SNe Ia.  Such an identification requires knowledge of the observable signatures of surviving companion stars of SNe Ia in the DDet scenario. 

Hypervelocity stars (HVSs) are generally defined as stars that move sufficiently fast to escape the gravitational potential of our Galaxy, i.e., they typically have velocities larger than $\sim400\,\rm{km\,s^{-1}}$ in the Galactic rest frame (for a review see \citealt{Brown2015a}).  HVSs are believed to be ejected by three-body interactions with the supermassive black hole at the Galactic center (e.g., \citealt{Hill88, Yu03, Brow05}), exchange encounters in other dense stellar environments (e.g., \citealt{Aars74}) between hard binaries and massive stars (e.g., \citealt{Leon91, Gvar09}), or disruptions of close binaries via SN explosions (e.g., \citealt{Blaa61, Taur98, Zubo13, Taur15, Geie15,Neunteufel2021}).  

The hypervelocity star US~708 has been classified as a sdO/B star. Based on a spectroscopic and kinematic analysis, \citet{Geie15} have reported that US~708 travels with a velocity of about $1\mathord,200\,\rm{km\,s^{-1}}$, suggesting that it is one of the fastest unbound stars in our Galaxy that was ejected from the Galactic disc $14.0\pm3.1\,\mathrm{Myr}$ ago. Considering the possibilities of different acceleration mechanisms of HVSs, \citet{Geie15} further concluded that US~708 is very unlikely to originate from the Galactic center, but it is rather the ejected donor remnant of a DDet SN~Ia.

\citet{Geie15}, however, did not compare the long-term evolution and appearance of surviving companion stars of DDet SNe Ia with the observations of US~708. It is therefore still an open question whether the observations of this star match the expected properties of a surviving SN Ia donor star in the DDet scenario.  The present paper predicts observable signatures of the surviving companion stars of sub-Chandrasekhar mass double-detonation SNe Ia by combining the outcomes of three-dimensional (3D) hydrodynamic simulations of ejecta-companion interaction with detailed one-dimensional (1D) calculations of the long-term evolution of the donor remnant.  The results of our models are then compared to the observations of US~708 to assess its proposed origin from a SN Ia ejection.

\begin{figure*}
  \begin{center}
    {\includegraphics[width=\textwidth, angle=0]{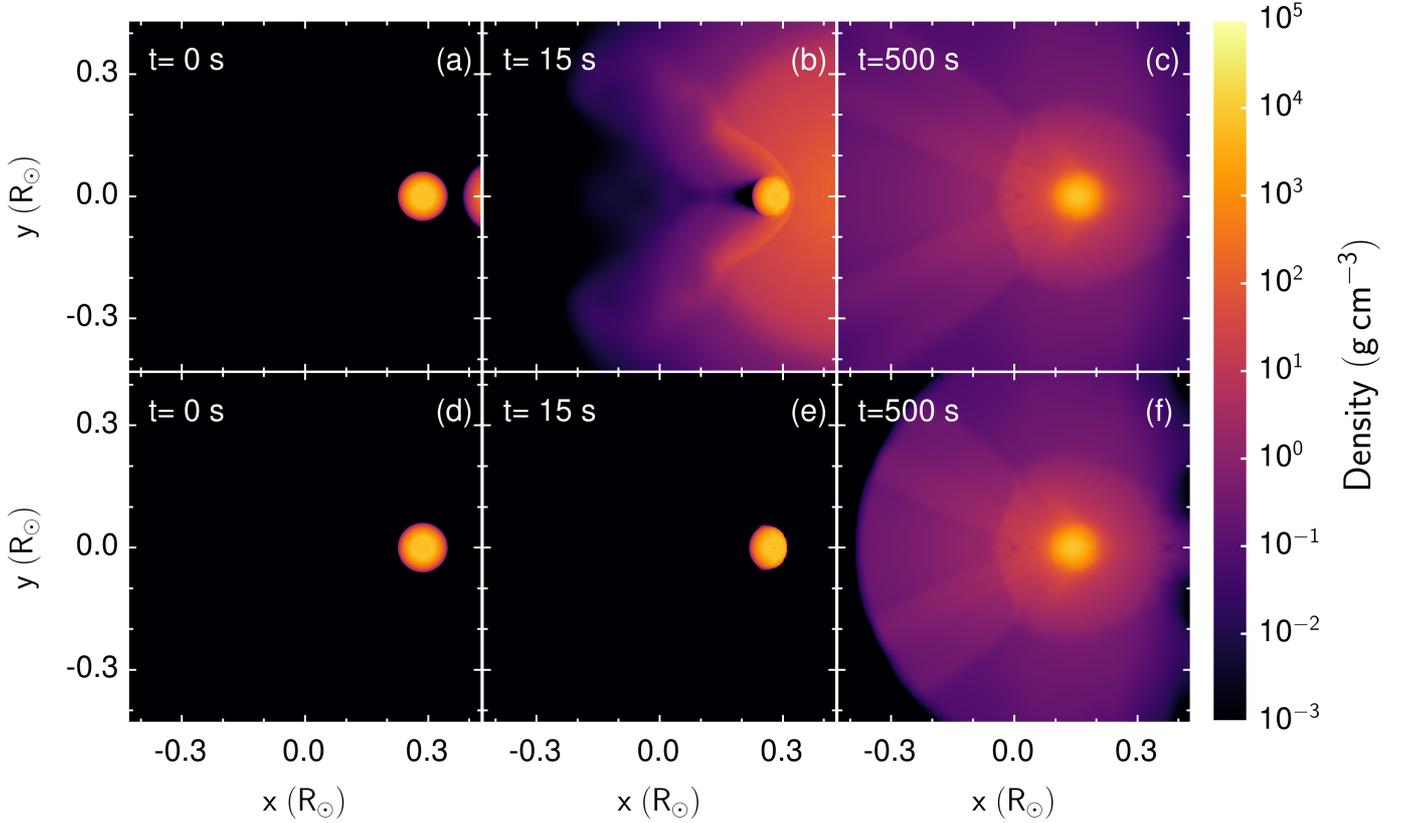}}
  \caption{Density distributions of all material (\textsl{Panels a--c}) and of the bound companion material (\textsl{Panels d--f}).  The panels shows a slice in the orbital plane in for three different times each. \textsl{Colours} indicate density as per colour bar (\textsl{right}). The 3D hydrodynamical simulation of the SN~ejecta-companion interaction shown here does not include binary orbital motion and stellar spin. All panels use the same length scales.}

\label{Fig:rho}
  \end{center}
\end{figure*}

\begin{figure}
  \begin{center}
    {\includegraphics[width=\columnwidth, angle=0]{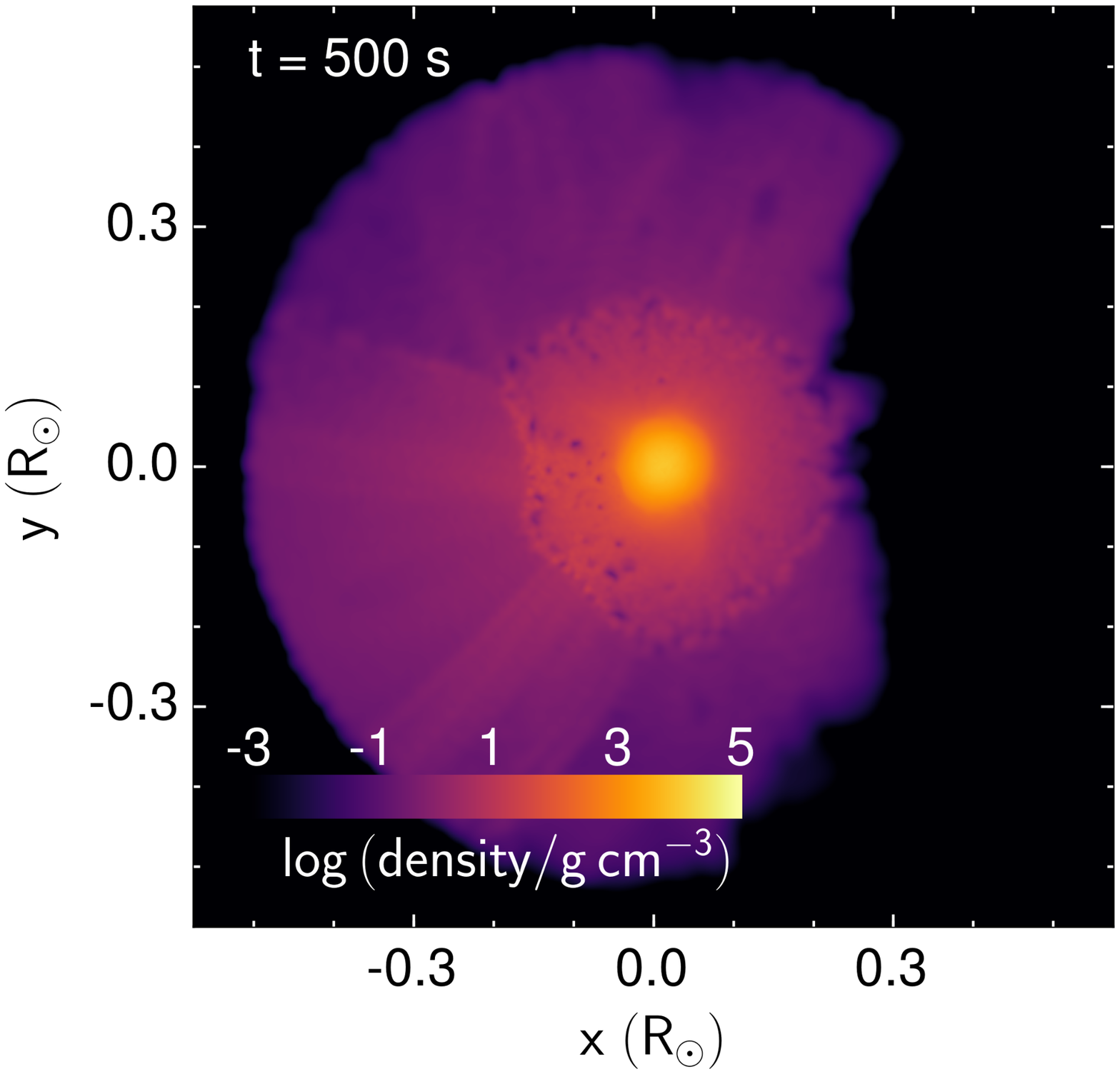}}
  \caption{Similar to Fig.~\ref{Fig:rho}f, but for the 3D simulation that includes binary orbital motion and stellar spin.}
\label{Fig:remnant}
  \end{center}
\end{figure}

The methods, the companion models and the SN Ia explosion model used in this study are described in Section~\ref{sec:code}. In Section~\ref{sec:results} we present the results of 3D simulations of ejecta-donor interaction and 1D post-explosion evolution calculations of the surviving companion stars. A comparison with the observation of US~708 and some discussion of our models is given in Section~\ref{sec:discussion}. The main conclusions are summarized in Section~\ref{sec:summary}.

\section{Numerical methods and models}
\label{sec:code}

We construct the initial He star companion models at the moment of SN Ia explosion based on a SN explosion model within the sub-Chandrasekhar mass DDet scenario of \citet{Gronow2020}.  We describe in detail the initial conditions and setup for our 3D hydrodynamic impact simulation, the conversion between the 3D Smooth Particle Hydrodynamics (SPH) models and the 1D model, as well as the 1D post-impact evolution of a surviving companion star.

\subsection{Code for the ejecta-donor interaction}

To obtain the detailed post-explosion properties of surviving companion stars of SNe Ia, we perform a 3D hydrodynamic simulation of SN ejecta-companion interaction with the SPH code {\sc Stellar GADGET} \citep{Spri05, Pakmor2012a}.  The initial conditions and basic setup for the impact simulation of this work are similar to those in our previous impact studies \citep[e.g.,][]{Liu2012,Liu2013a,Liu2013b,Liu2013c}. The companion star at the moment of SN explosion and the SN ejecta model are the two fundamental inputs of our hydrodynamic impact simulation, and we discuss our setups for these two models below.

\begin{figure}
  \begin{center}
    {\includegraphics[width=\columnwidth, angle=0]{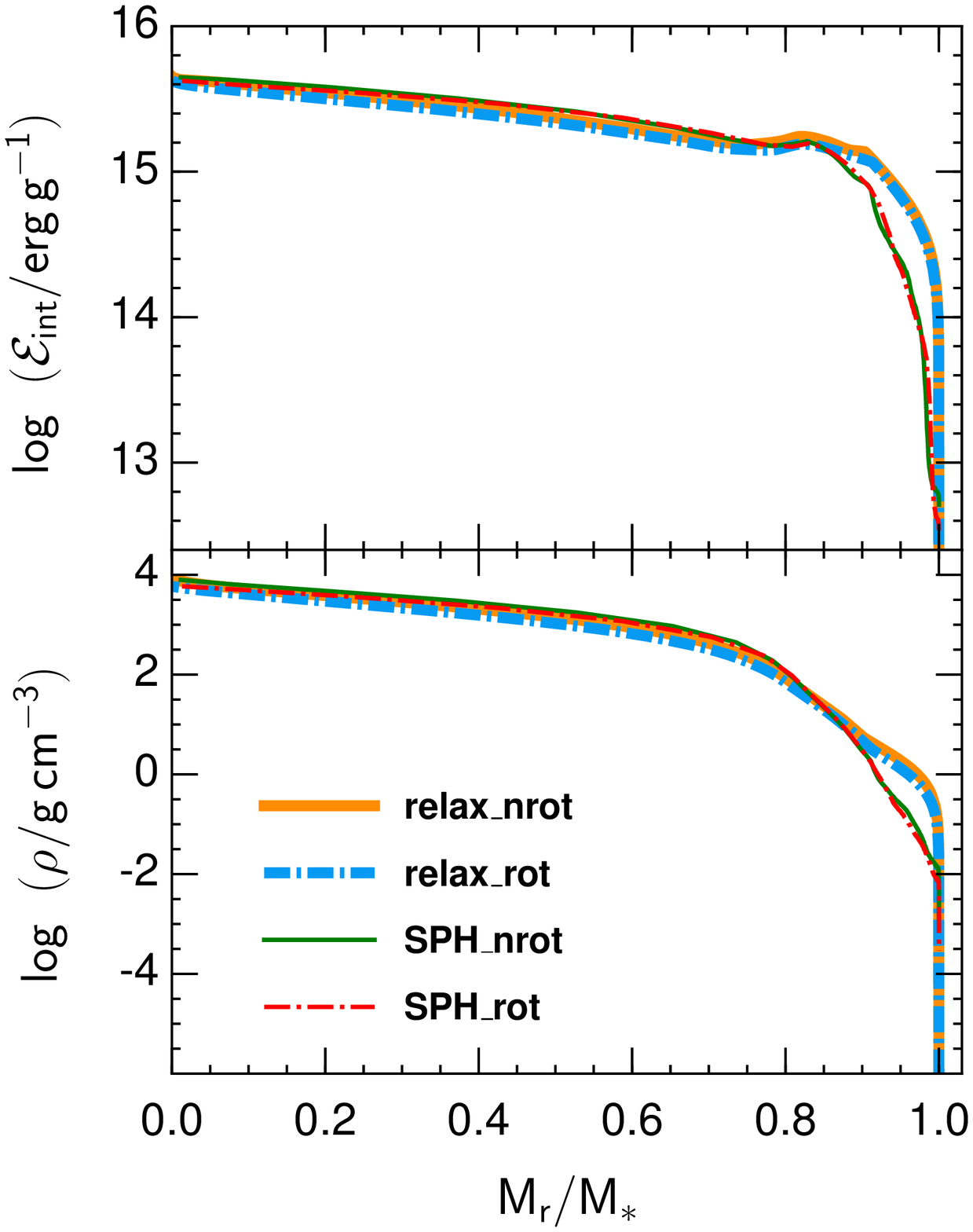}}
    \caption{Post-impact 1D angle-averaged profiles (\textsl{thin lines}) of specific internal energy $\mathcal{E}_{\mathrm{int}}$ (\textsl{Top Panel}) and density $\rho$ (\textsl{Bottom Panel}) as functions of fractional mass coordinate at the end of the SPH impact simulation for our reference Model~\texttt{A}.  For a comparison, the relaxed starting model in \textsc{MESA} are shown as  \textsl{thick lines}.  The results for the impact simulation with (or without) binary orbital motion and stellar spin taken into account are shown as \textsl{dash-dotted lines} and \textsl{solid lines}, respectively.}
\label{Fig:profile}
  \end{center}
\end{figure}

\begin{figure*}
  \begin{center}
    {\includegraphics[width=\columnwidth, angle=0]{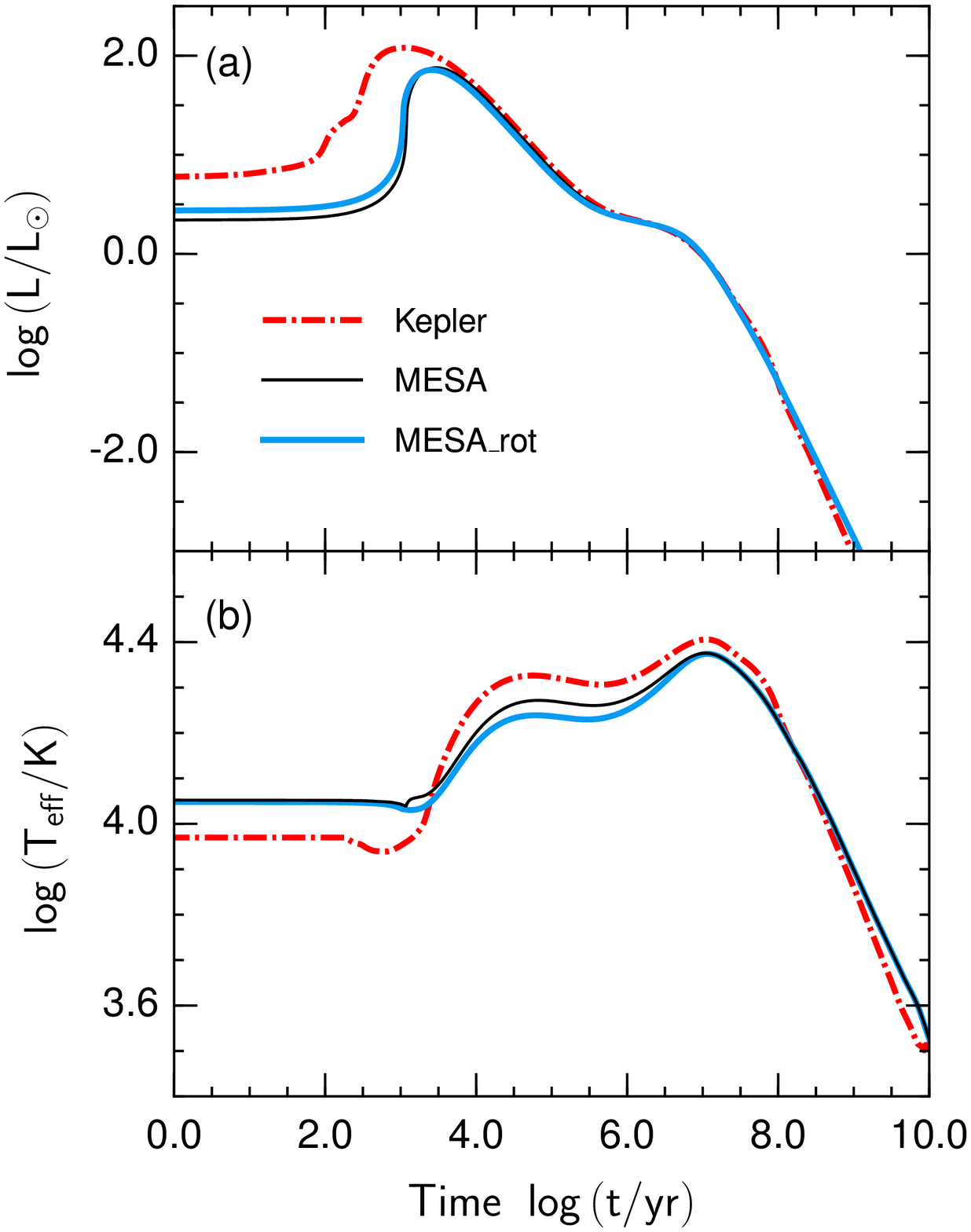}}
    \hfill
    {\includegraphics[width=\columnwidth, angle=0]{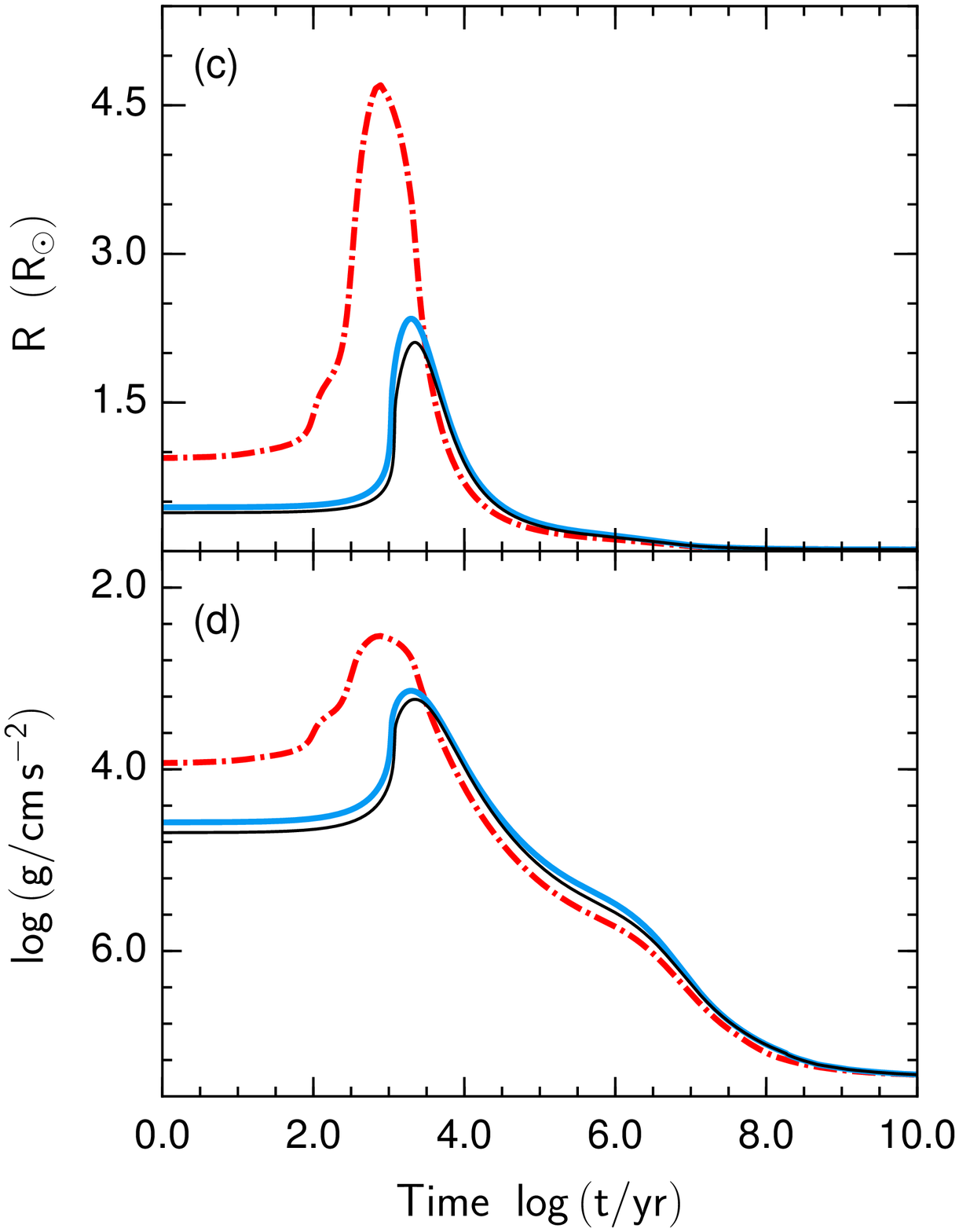}}
  \caption{Post-impact evolution of the photosphere luminosity, $L$, effective temperature, $T_{\mathrm{eff}}$, radius, $R$, and surface gravity, $g$, of a surviving He star companion as functions of time.  The \textsl{solid line} and \textsl{dash-dotted line} respectively correspond to the results from 1D calculations using the \textsc{Kepler} code and by the \textsc{MESA} code.  For comparison, the \textsc{MESA} results based on the impact simulations that do and do not include the binary orbital motion and stellar spin are given as \textsl{thick solid line} and  \textsl{thin solid line}, respectively.}
\label{Fig:evolution}
  \end{center}
\end{figure*}

\begin{figure*}
  \begin{center}
    {\includegraphics[width=\columnwidth, angle=0]{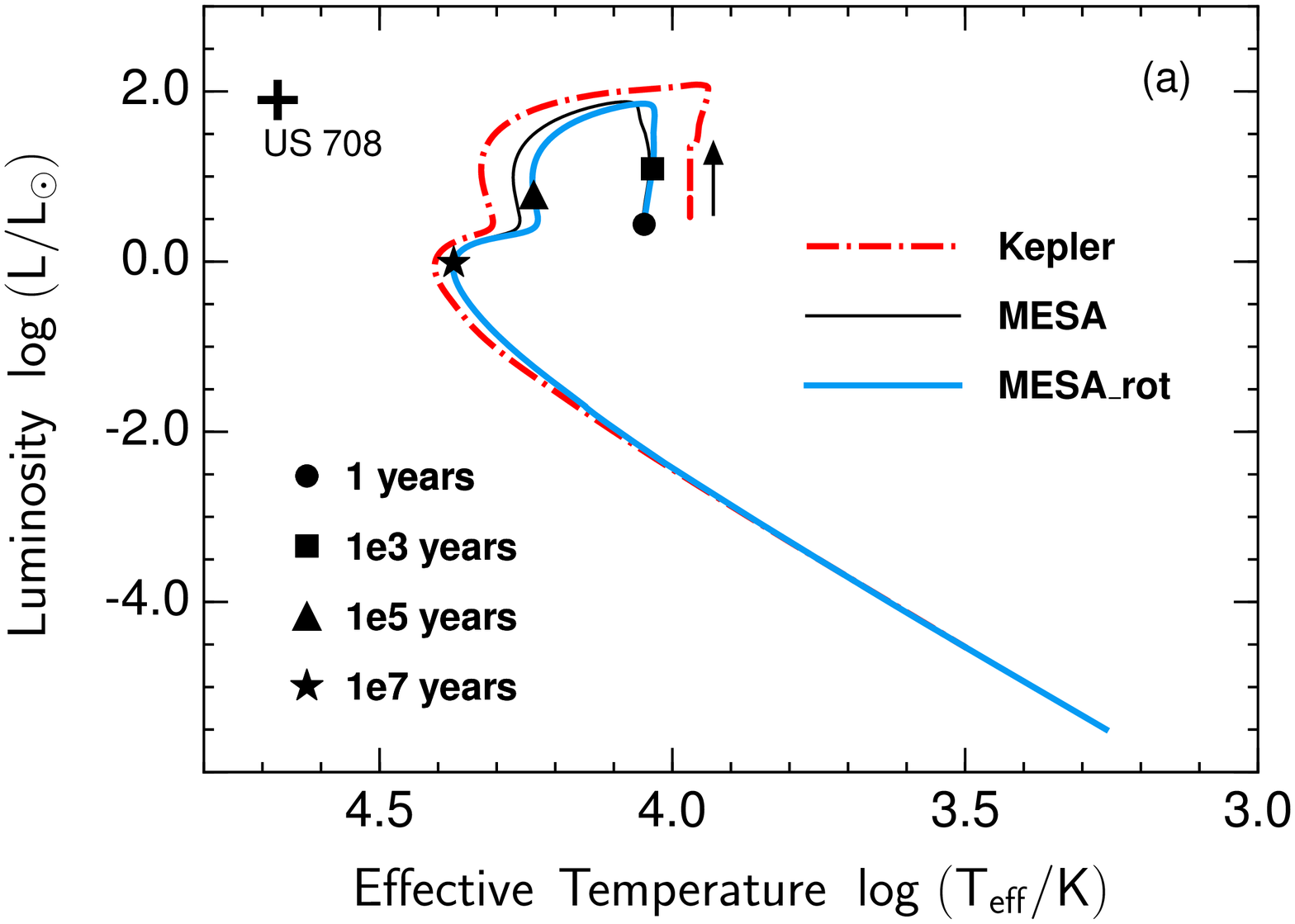}}
    \hfill
    {\includegraphics[width=\columnwidth, angle=0]{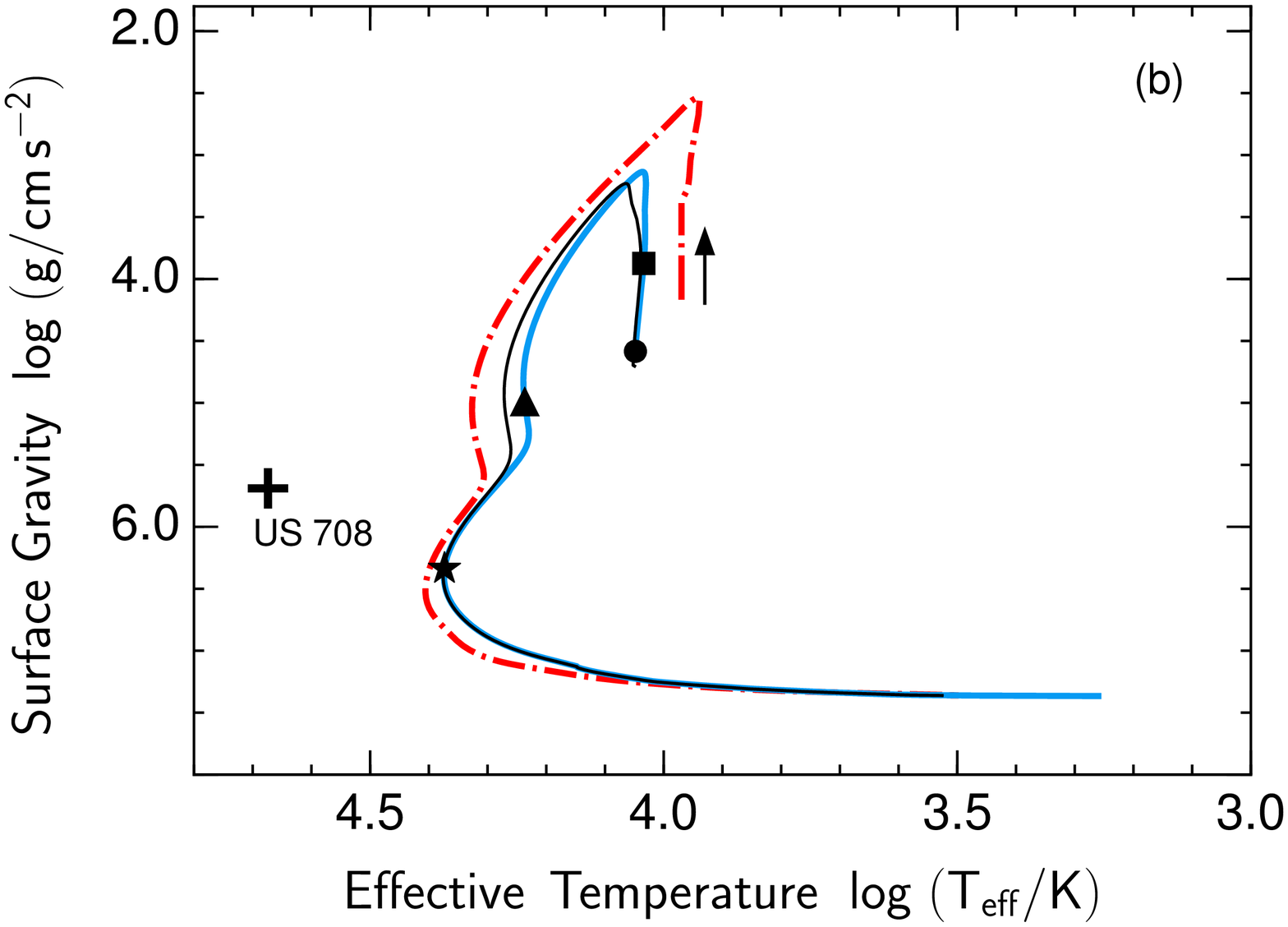}}

    \caption{Post-impact evolutionary tracks of a surviving He star companion model in the Hertzsprung-Russell Diagram (\textsl{Left Panel}) and surface gravity vs.\ temperature diagram (\textsl{Right Panel}).  The \textsl{solid lines} and \textsl{dash-dotted lines} respectively correspond to the results of \textsc{Kepler} and \textsc{MESA} calculations. The \textsl{filled circle}, \textsl{square}, \textsl{triangle}, and \textsl{star markers} on the tracks present post-impact evolutionary phases of $1\,\mathrm{yr}$, $1\,\mathrm{kyr}$, $100\,\mathrm{kyr}$, and $10\,\mathrm{Myr}$ after the SN impact, respectively. The \textsl{thin lines} and \textsl{thick lines} have the same meaning as those in Fig.~\ref{Fig:evolution}.}

\label{Fig:track}
  \end{center}
\end{figure*}

\subsubsection{The He-star donor model}
\label{sec:donor}

\citet{Geie15} proposed a double detonation SN explosion resulting from a binary progenitor system composed of a WD with $M_{\mathrm{WD}}^{\mathrm{i}}=1.05\,{\rm M}_{\odot}$ WD and a $M_{2}^{\mathrm{i}}=0.45\,{\mathrm M}_{\odot}$ He-burning companion star with an initial orbital period of $P^{\mathrm{i}}\simeq 0.0182$ days as a possible origin of US~708.  In their model, the first detonation is assumed to be triggered when the mass of the accreted He shell reaches $M_{\mathrm{He}}=0.15\,{\rm M}_{\odot}$.  The surviving donor He star of $M_{2}^{\mathrm{f}}=0.30\,{\rm M}_{\odot}$ is ejected from the system.  In this model, the He donor has an orbital velocity of $V_{\mathrm{orb}}\simeq 920\,\mathrm{km\,s^{-1}}$ at the moment of SN Ia explosion.  This can explain the observed high velocity of US~708 ($1200\,\mathrm{km\,s^{-1}}$; \citealt{Geie15}). \citet{Woosley2011, Bildsten2007, Shen2009, Shen2014, Shen2018}, for example, suggest that a massive WD could trigger a He detonation by accreting a rather thin He shell from its companion star.  For this scenario, recent 2D/3D hydrodynamic models of SN Ia explosion have also shown that a core detonation is still possible despite the weaker He shell detonation \citep{Fink10, Sim10, Townsley2019, Gronow2020, Gronow2021}.  This seems to indicate that a massive initial WD of $\mathord\sim1.05\,{\rm M}_{\odot}$ is likely to accumulate a thin He shell rather than the $0.15\,{\rm M}_{\odot}$ required otherwise to successfully trigger a thermonuclear explosion by undergoing a double detonation.

\begin{figure*}
  \begin{center}
    {\includegraphics[width=\columnwidth, angle=0]{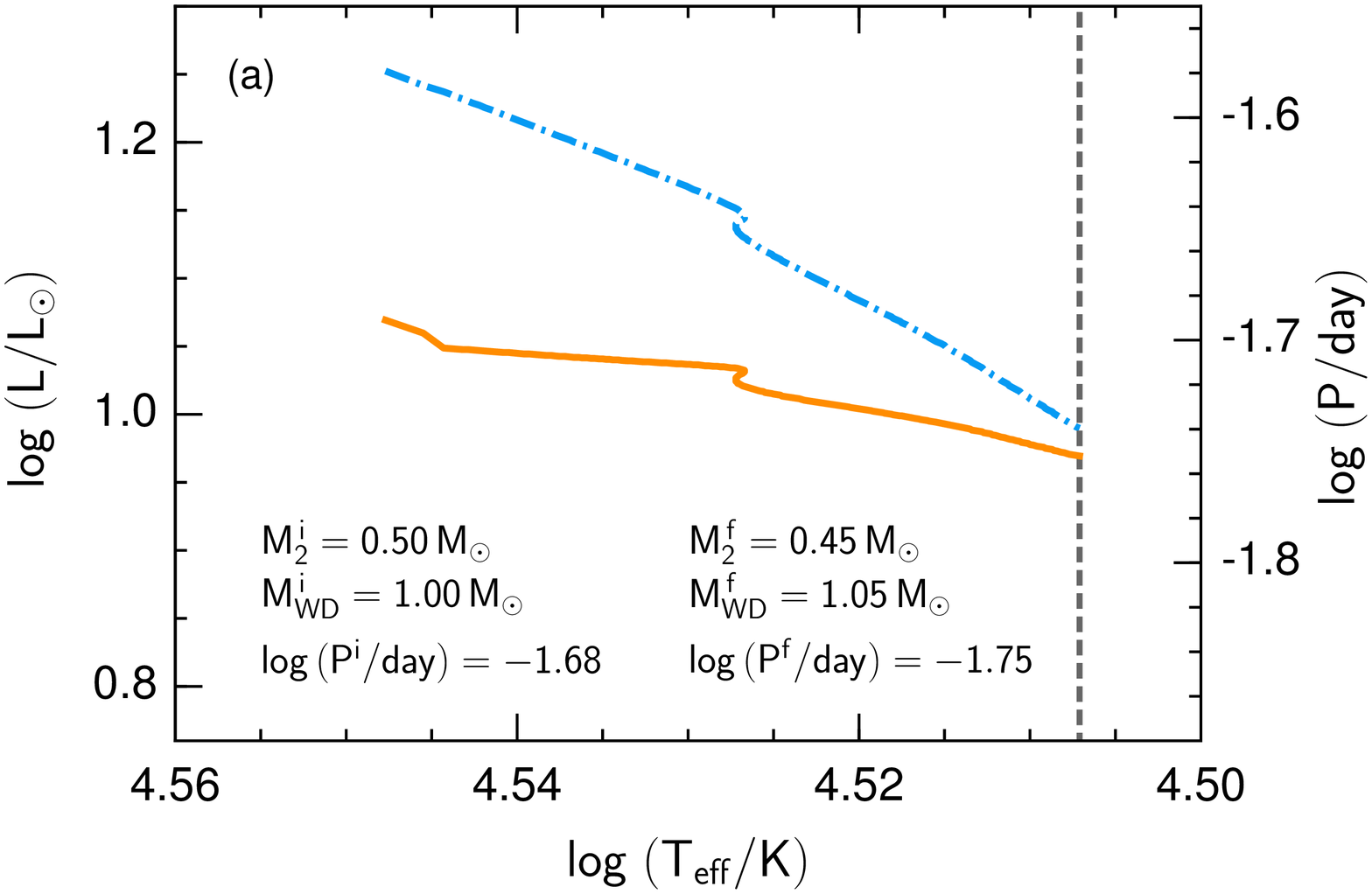}}
    \hfill
    {\includegraphics[width=\columnwidth, angle=0]{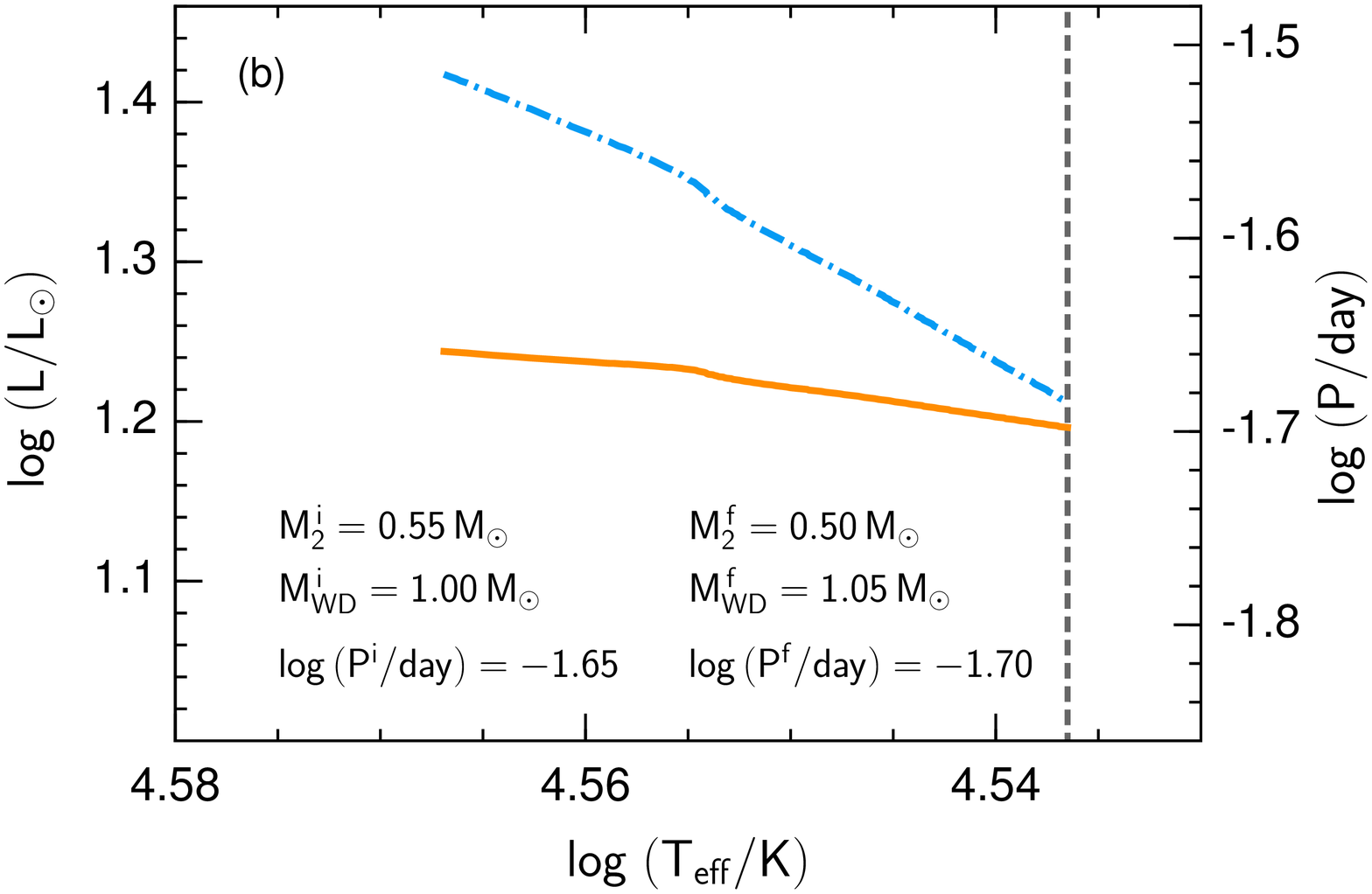}}
    \caption{Similar to the \textsl{Left Panel} of Fig.~\ref{Fig:star1}, but for two more massive He-star donor models with initial masses of $0.50\,{\rm M}_{\odot}$ (\textsl{Left Panel}) and $0.55\,{\rm M}_{\odot}$ (\textsl{Right Panel}).}

\label{Fig:star2}
  \end{center}
\end{figure*}

We adopt the method of \citet{Geie15} to construct the He-star donor model at the moment of explosion by performing binary evolution calculation using the 1D stellar evolution code \textsc{MESA} \citep{Paxton2011, Paxton2015, Paxton2018}\footnote{\citet{Geie15} used the Cambridge stellar evolution code \textsc{STARS} \citep{Eggl71,Eggl72,Pols1995,Stancliffe2010} for their binary evolution calculations.  When a consistent setup is adopted, we find that the donor structures and binary parameters are identical between the \textsc{MESA} and the \textsc{STARS} models at the moment of SN explosion.}.  As detailed in  Sect.~\ref{sec:sn}, our goal is to directly connect to the DDet Model \texttt{M2a}\footnote{Model \texttt{M2a} was constructed to be similar to the Model~\texttt{3} of \citet{Fink10}, and it was presented as the reference model in \citet{Gronow2020}.} of \citet{Gronow2020}.  We therefore diverge from the system parameters given by \citet{Geie15} and start our consistent binary evolution calculation with a progenitor system consisting of an initial WD ($M_{\mathrm{WD}}^{\mathrm{i}}=1.00\,{\rm M}_{\odot}$) and a He-burning star ($M_{2}^{\mathrm{i}}=0.35\,{\rm M}_{\odot}$) with an orbital period of $P^{\mathrm{i}}\simeq 0.0135$ days (Table~\ref{table:1}).  Figure~\ref{Fig:star1} shows key parameters of the detailed binary evolution calculation of the He-star donor model used for our reference impact simulation.  The initial He companion model is set up with a He abundance of $Y=0.98$, with a metallicity of $Z=0.02$, and the WD is treated as a point mass in our binary evolution calculation.  This binary system starts mass transfer when the He donor star fills its Roche lobe.  The orbital angular momentum loss due to gravitational wave radiation is included by following the standard formula given by \citet{Landau1971}. According to previous works, at a mass transfer rate of $\dot{\mathrm{M}}_{\mathrm{tr}} < 4.0 \times 10^{-8}\,\mathrm{M}_{\odot}\,\mathrm{yr}^{-1}$ the He shell builds up steadily, avoiding He burning \citep[e.g.,][]{Woosley1986, Ruiter2014}.  Matching to the hydrodynamic explosion model \texttt{M2a} of \citet[][see their Table~1]{Gronow2020}, the double-detonation is assumed to trigger when the He-shell mass reaches a critical value of $M_{\mathrm{He}}$=$0.05\,\mathrm{M}_{\odot}$ \citep[see also][]{Fink10}. 

For our binary evolution simulation we obtain a He donor model with a mass of $M_{2}^{\mathrm{f}}\approx 0.30\,{\rm M}_{\odot}$ (Fig.~\ref{Fig:star1}) at the moment of SN Ia explosion.  This model is used as input of our 3D supernova explosion impact simulation.  This is our reference model (Table~\ref{table:1}).  The properties of this companion model, its density profile, the orbital velocity of $V_{\mathrm{orb}}\simeq 901\,\mathrm{km\,s^{-1}}$, and the effective temperature of $\mathrm{log_{10}}\,T_{\mathrm{eff}}\simeq 4.1$, are quite similar to those suggested by \citet{Geie15} for US~708.

\subsubsection{The SN explosion model}
\label{sec:sn}

Our binary progenitor models are constructed to match the hydrodynamic simulation Model \texttt{M2a} of \citet{Gronow2020} for a double-detonation sub-Chandrasekhar mass SN~Ia explosion. We directly adopt their Model \texttt{M2a} to represent SN Ia explosion here.  This model has masses of the exploding star, of the C/O core, and of the He shell that are $M_{\mathrm{tot}}\approx 1.05\,{\rm M}_{\odot}$, $M_{\mathrm{core}}\approx 1.00\,{\rm M}_{\odot}$, and $M_{\mathrm{shell}}\approx 0.05\,{\rm M}_{\odot}$, respectively \citep[][see their Table~1]{Gronow2020}.  The total explosion energy is about $1.25\times 10^{51}\,\mathrm{erg}$, and the average velocity of the ejecta is of the order $10^{4}\,\mathrm{km\,s^{-1}}$. \citet{Gronow2020} and \citet{Fink10} provide a detailed description of this explosion model. In addition, \citet{Gronow2020} present the results of a time-dependent multi-wavelength radiative transfer calculations for this model.

\subsubsection{Initial setup}

As described in \citet{Pakmor2012a}, we use the HEALPix method \citep[][]{Gorski2005} to map the 1D profiles of density and internal energy of our He-star donor model to a particle distribution suitable for the 3D SPH code \citep[see also][]{Liu2012,Liu2013a,Liu2013b,Liu2013c}.  Before we start the actual impact simulation, the SPH model of each donor star is relaxed for ten dynamical timescales, $t_{\rm{dyn}}=\nicefrac12 \,(G\rho)^{-\nicefrac12}$, to reduce numerical noise introduced by the mapping \citep[e.g.,][]{Pakmor2012a,Liu2012}.

For our reference simulation, about $1.8\times10^7$ SPH particles are used in total to represent both the He companion star and  the explosion model.  All SPH particles are assigned the same mass of about $7.4\times10^{-8}\,\mathrm{M}_{\odot}$.  Once the relaxation of the He-star donor is finished, the double-detonation sub-Chandrasekhar explosion model described above is placed at the distance to the donor star that given by the binary separation at the moment of SN Ia initiation in our 1D binary evolution calculation.  Based on the 1D-averaged radial profiles of Model \texttt{M2a} of \citet{Gronow2020}, SPH particles are placed randomly in shells to reproduce the density profile.  The chemical composition and radial velocity of each particle are set to the values of the initial 1D explosion model at a given radius.  The ejecta-companion interaction is then simulated for several thousands of seconds until it has ceased and the total stripped mass and kick velocity received by the donor due to the interaction have saturated.

For our second set of impact simulations that take into account the orbital and stellar spin velocities of the companion star, we assume that the companion star co-rotates with its orbit due to strong tidal interaction during pre-explosion mass-transfer phase. This leads to locking of the spin period of the companion star with its orbital period.  The companion star is assumed to rotate as a solid body.  To isolate the impact of the binary model parameters, we use the same explosion model described in Section~\ref{sec:sn} in all simulations presented of this work.

\subsection{Mapping from 3D to 1D}
\label{sec:conversion}

Predicting the observable properties of a surviving companion star in the supernova remnant phase requires  to model its post-impact evolution over a few hundred years after the SN Ia explosion. Since the time step in our 3D SPH impact simulations is on the order of the dynamical timescale -- hundreds of seconds -- this is not possible in
the framework of our 3D hydrodynamic simulation.  To limit the computational cost of the simulation to a reasonable wall clock time, the outcome of 3D impact simulations is mapped into the 1D stellar evolution code \textsc{MESA} by using the method of \citet[][see their Section~2]{Liu2021} to trace the subsequent long-term post-explosion evolution \citep[see also][]{Pan2013, Pan2014,Bauer2019}.  The three main steps of this method are briefly summarized in the following:  First, the 3D post-impact companion models from our hydrodynamic impact simulations (Fig.~\ref{Fig:rho}f) are converted into 100--200 spherical shells.  The physical properties of the SPH particles, such as the internal energy and composition, are averaged to give a value for that shell.  Second, the 1D averaged radial profiles of internal energy, chemical composition, and density are used as inputs for the \textsc{MESA} code to compute suitable starting models for the subsequent post-explosion calculations by directly adopting the relaxation routines provided in \textsc{MESA} \citep[][see their Appendix B]{Paxton2018}.  Third, we follow long-term evolution of a surviving companion model to predict its observable properties until it enters the WD cooling phase.

We use the MESA code in its hydrostatic mode and therefore artificially relax our hydrodynamic post-impact model to hydrostatic equilibrium.  To test the effect of this approximation, we additionally perform a simulation of the post-impact evolution for our reference model with the 1D hydrodynamic stellar evolution code \textsc{Kepler} \citep{Weaver1978,Rauscher2002,Woosley2002,Heger2010}.  Similarly, the angle-averaged 1D radial profiles obtained from 3D post-impact companion models are used as inputs of the subsequent \textsc{Kepler} calculations, but without performing any relaxation process.  In addition, we include the radial velocity of each spherical shell at the end of our impact simulation into the \textsc{Kepler} models.  This way the hydrodynamics is also followed from the beginning of of the mapping in the \textsc{Kepler} calculations.  We find no significant difference between \textsc{MESA} and \textsc{Kepler} results for our reference calculation for Model~\texttt{A} (Section~\ref{sec:1d}).  For consistent analysis and discussion, all other He-star donor models in Section~\ref{sec:models} are only simulated using the \textsc{MESA} code.

\begin{table*}\renewcommand{\arraystretch}{1.2}
\fontsize{8}{12}\selectfont
\caption{Three He donor models studied by this work.} \label{table:1}
\centering
\begin{tabular}{ccccccccccccccc}    
\hline\hline
Model   & $M^{\mathrm{\,i}}_{\mathrm{WD}}$  & $M^{\mathrm{\,i}}_{\mathrm{2}}$  & $M^{\mathrm{\,f}}_{\mathrm{2}}$  &$\mathrm{log}\,P^{\mathrm{f}}$ & $\mathrm{log}\,T_{\mathrm{eff}}^{\mathrm{f}}$ & $a^{\mathrm{f}}$ & $R^{\mathrm{f}}_{\mathrm{2}}$   & $V^{\mathrm{\,f}}_{\mathrm{orb}}$ & $V^{\mathrm{\,f}}_{\mathrm{rot}}$ & $\Delta M/M^{\mathrm{\,f}}_{2}$ & \ \ $V_{\rm{kick}}$    & $L_{\mathrm{peak}}$  & $t_{\mathrm{peak}}$ & $E_{\mathrm{in}}$ \\
        &($\mathrm{M}_{\rm{\odot}}$)                 &($\mathrm{M}_{\rm{\odot}}$)               & ($\mathrm{M}_{\rm{\odot}}$)               &(days)                               & ($K$)   & \multicolumn{2}{c}{($10^{10}\,\mathrm{cm}$)}& ($\mathrm{km\,s^{-1}}$) & ($\mathrm{km\,s^{-1}}$) & (\%)  &($\mathrm{km\,s^{-1}}$)& ($10^{3}\,\mathrm{L}_{\odot}$)  & ($\mathrm{10^{3}\,yr}$) &($10^{49}\,\mathrm{erg}$)\\
\hline
   A & 1.00 & 0.35 & 0.30 & -2.08 & 4.10 & 1.34 & 0.37 & 901 & 320 & 5.7 & 201 & 0.08 & 2.96 & 1.15\\
   B & 1.00 & 0.50 & 0.45 & -1.75 & 4.50 & 2.28 & 0.70 & 654 & 288 & 3.1 & 95  & 1.48 & 0.06 & 1.07\\
   C & 1.00 & 0.55 & 0.50 &- 1.70 & 4.53 & 2.50 & 0.77 & 614 & 287 & 2.8 & 83  & 2.09 & 0.04 & 1.08\\
\hline
\end{tabular}
\tablefoot{$M^{\mathrm{i}}_{\mathrm{WD}}$ and $M^{\mathrm{i}}_{\mathrm{2}}$ are the initial masses of the WD and donor at the beginning of mass transfer. $M^{\mathrm{\,f}}_{2}$, $P^{\mathrm{f}}$, $T_{\mathrm{eff}}^{\mathrm{f}}$, $a^{\rm{f}}$, $R^{\mathrm{f}}_2$, $V^{\mathrm{\,f}}_{\mathrm{orb}}$, and $V^{\mathrm{\,f}}_{\mathrm{rot}}$ denote the final mass, orbital period, the effective temperature, the binary separation, the radius, the orbital velocity, and the rotational velocity of donor star at the moment of SN explosion, respectively.  We assume that the He donor star is tidally locked to its orbital motion.  $\Delta M$ and $V_{\rm{kick}}$ are the total stripped donor masses and kick velocity caused by SN impact.  $L_{\mathrm{peak}}$ and $t_{\mathrm{peak}}$ give the peak luminosity and the corresponding post-impact evolutionary time of our \textsc{MESA} models before the deposited energy is radiated away, i.e., before the stars go back to the thermal equilibrium state. $E_{\mathrm{in}}$ corresponds to the total amount of energy deposited into the donor star due to the interaction with SN Ia ejecta.
}

\end{table*}

\section{Numerical results}
\label{sec:results}

In this section, we present the results of 3D hydrodynamic simulations of ejecta-companion interaction. The 1D post-explosion evolution calculations of the surviving companion stars with the \textsc{MESA} and \textsc{Kepler} code are also given.

\subsection{Ejecta-donor interaction}
\label{sec:impact}

Figure~\ref{Fig:rho} shows the density distribution of all material (top row) and bound donor material (bottom row) as a function of time in our 3D impact simulation for our reference He-star donor model (Model~\texttt{A}).  The figure is oriented such that the SN explodes at right-hand side of the donor star.  After the explosion, the SN ejecta expand freely for a while (a few 10 seconds) and hits the surface of donor star, stripping some He-rich material from its surface and forming a bow shock. Subsequently, the bow shock propagates through the donor star (Fig.~\ref{Fig:rho}b), removing more He-rich material from the far side of it.  The donor star is significantly shocked and heated during the interaction with the supernova blast wave.  It inflates dramatically, but survives the interaction and starts to relax (Fig.~\ref{Fig:rho}f). The stripped He-rich donor material is largely embedded in the low-velocity SN debris behind the star. Quantitative details of the models and of the simulation results are summarised in Table~\ref{table:1}.

At the end of our 3D simulation, we find that the total amount of stripped He material from the donor surface during the ejecta-donor interaction is about $\Delta {M}\approx 0.02\,\mathrm{M}_{\sun}$, i.e.,\ about $6\,\%$ of the total donor mass.  The donor star receives a kick velocity of $\upsilon_{\mathrm{kick}}\approx 200\,\mathrm{km\,s^{-1}}$, which moves its remnant to left by about $10^{10}\,\mathrm{cm}$ at $500\,\mathrm{s}$ after the explosion (Fig.~\ref{Fig:rho}f).  The impact simulations for our reference donor model are ran with or without including binary orbital motion and stellar spin to test their effect on the results. Since the orbital and spin velocities of the donor star are much lower than the typical expansion velocity of SN ejecta of our explosion model ($\sim10^{4}\,\mathrm{km\,s^{-1}}$ according to  \citealp[][]{Gronow2020}), only small differences (less than 5\%) in the total amount of stripped donor mass and in the kick velocity are observed when the binary orbital motion and stellar spin are included. This is consistent with previous works \citep[e.g.,][]{Liu2013a,Liu2013b,Pan2012}. The density distribution of the ejected donor remnant from our impact simulations that includes binary orbital motion and stellar spin is shown in Fig.~\ref{Fig:remnant}. More asymmetric features in the morphology are observed for this case compared with that shown in Fig.~\ref{Fig:rho}f.

\begin{figure*}
  \begin{center}
    {\includegraphics[width=\columnwidth, angle=0]{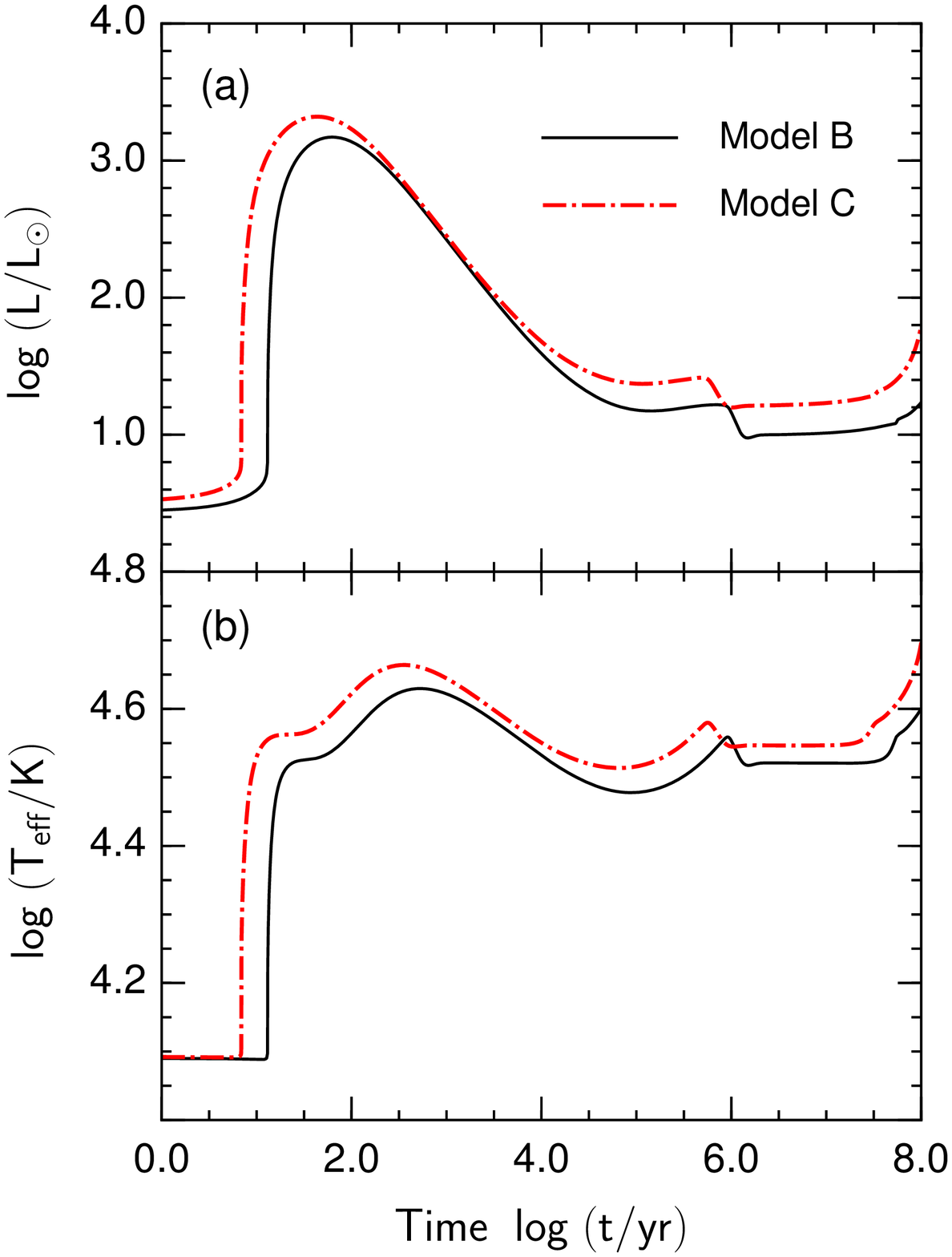}}
    \hfill
    {\includegraphics[width=\columnwidth, angle=0]{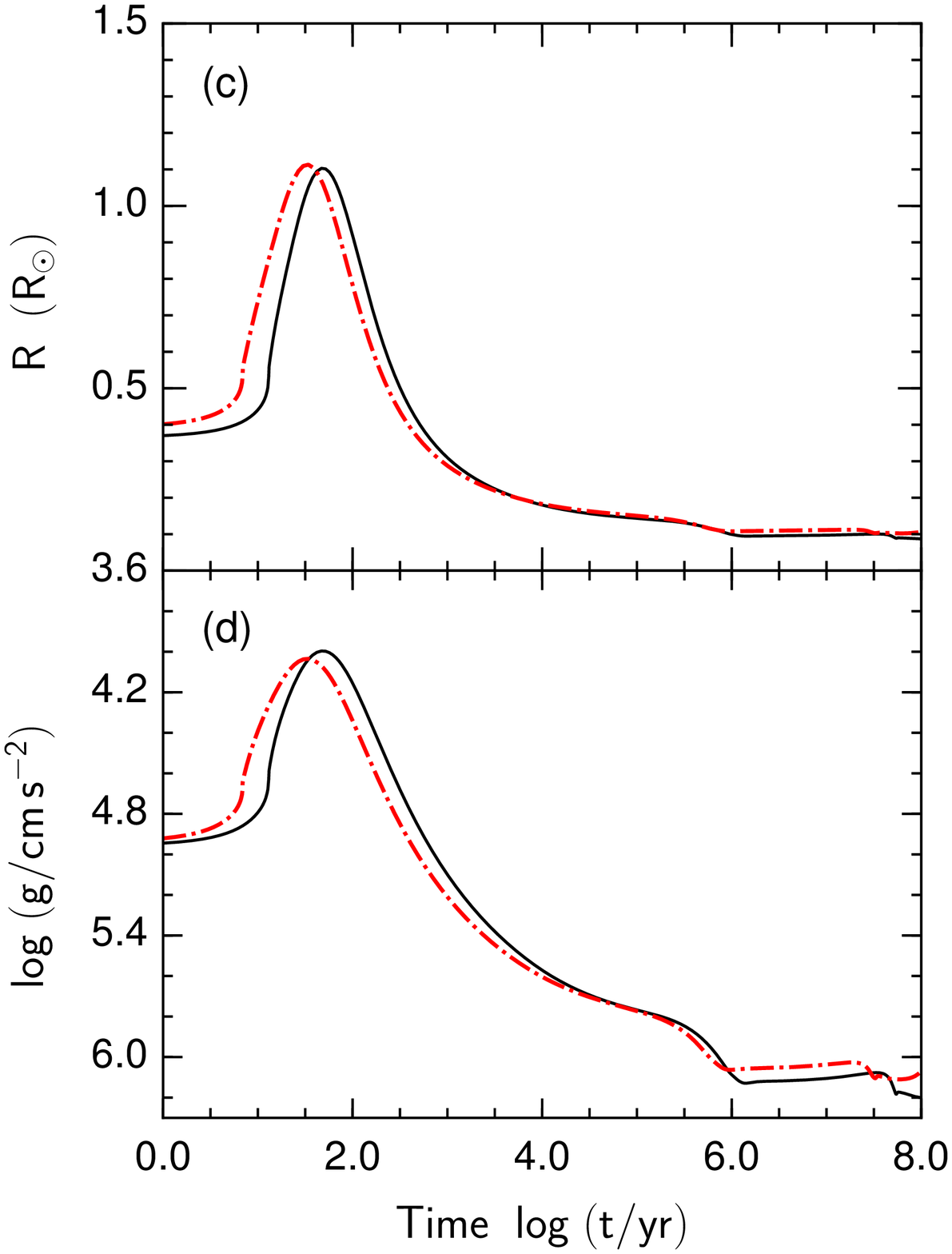}}

    \caption{Similar to Fig.~\ref{Fig:evolution}, but for the more massive He-star donor models with a post-impact mass of $0.45\,M_{\odot}$ (i.e., Model~\texttt{B}, see \textsl{solid black line}) and $0.50\,M_{\odot}$ (i.e., Model~\texttt{C}, see \textsl{red dash-dotted line}).}

\label{Fig:evolution2}
  \end{center}
\end{figure*}

\begin{figure*}
  \begin{center}
    {\includegraphics[width=\columnwidth, angle=0]{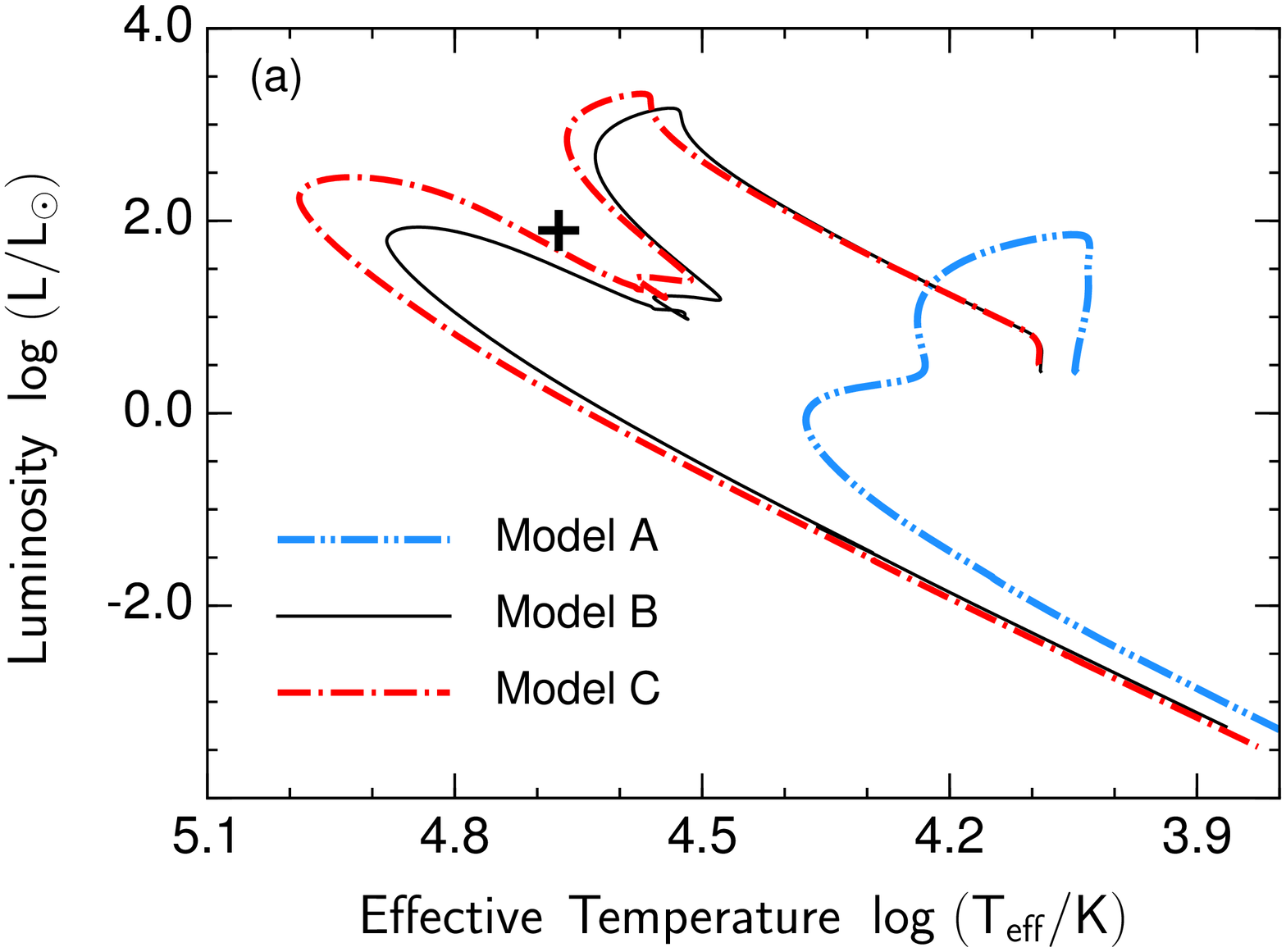}}
    \hfill
    {\includegraphics[width=\columnwidth, angle=0]{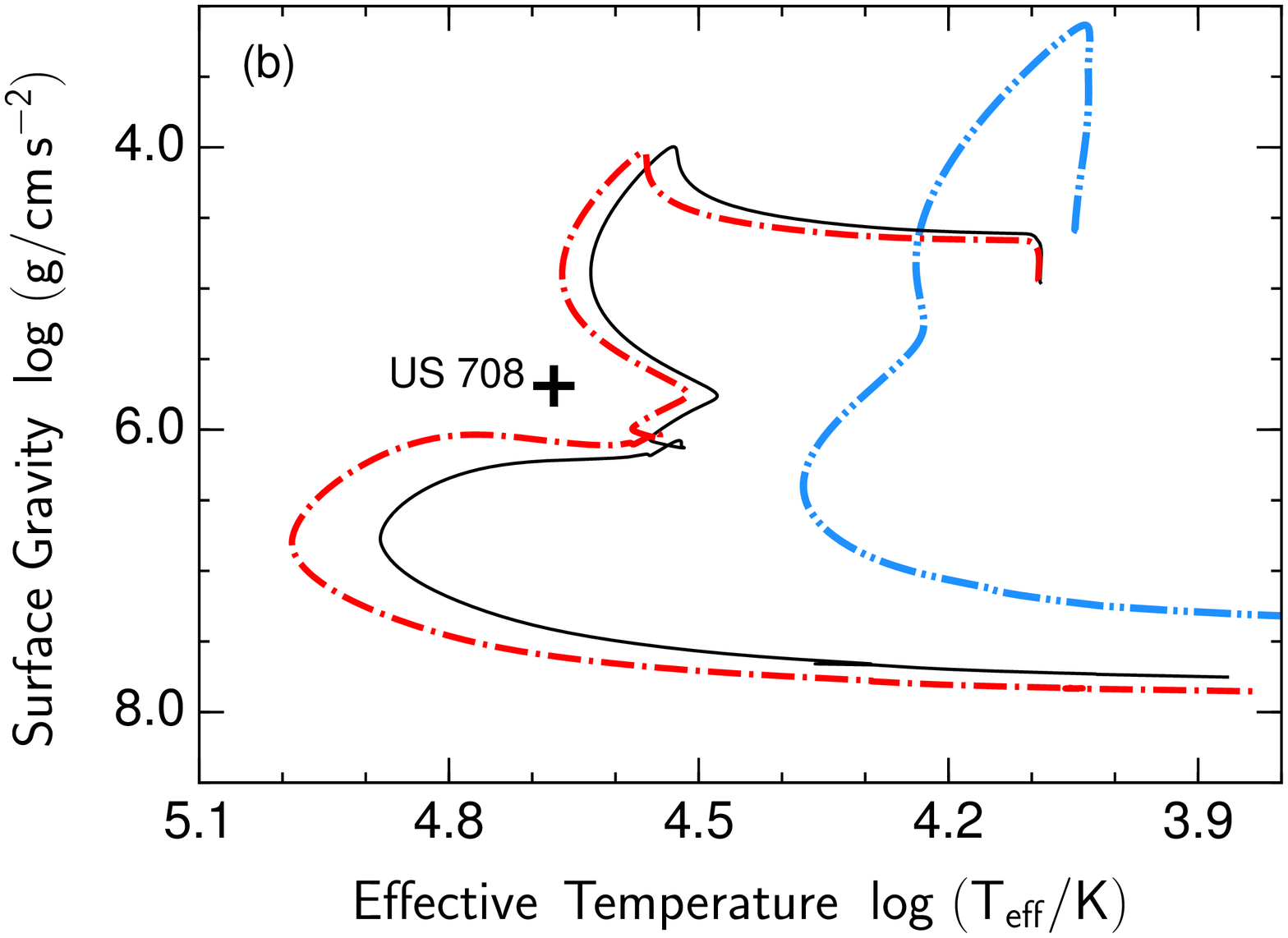}}

    \caption{Similar to Fig.~\ref{Fig:track}, but for the more massive He-star donor models with a post-impact mass of $0.45\,M_{\odot}$ (\textsl{solid black line}) and $0.50\,M_{\odot}$ (\textsl{red dash-dotted line}).  For a comparison, the results of $0.30\,M_{\odot}$ He donor model of Fig.~\ref{Fig:track} is shown as a \textsl{double-dotted line}.  The observed properties of US~708 are indicated as a \textsl{black plus} `\textbf{+}'.}

\label{Fig:track2}
  \end{center}
\end{figure*}

\subsection{1D post-impact evolution}
\label{sec:1d}

As described in Section~\ref{sec:conversion} \citep[see also][]{Liu2021}, the final outcomes of our 3D impact simulations (Fig.~\ref{Fig:remnant}) are used to construct the initial inputs for the \textsc{MESA} and \textsc{Kepler} codes to model the long-term post-impact evolution of the surviving donor stars.  Figure~\ref{Fig:profile} shows a comparison between 1D-averaged internal energy and density profiles of our reference model (i.e., Model~\texttt{A} in Table~\ref{table:1}) at the end of impact simulation and of the corresponding post-relaxed profiles in \textsc{MESA}.

Figure~\ref{Fig:evolution} shows the post-impact evolution of photospheric luminosity, $L$, the effective temperature, $T_{\mathrm{eff}}$, the radius, $R$, and the surface gravity, $g$, as functions of time for our reference model (Model~\texttt{A} in Table~\ref{table:1}).  The donor star is significantly shocked by the SN impact during the ejecta-donor interaction.  This leads to an energy deposition of $1.15\times10^{49}\,\mathrm{erg}$.   This value has been calculated by tracing the increase in binding energy of the star after the SN Ia impact \citep[see also][]{Pan2012b, Liu2021}.  We find that $\sim$$50\%$ of the incident energy of the SN Ia ejecta, about $2.30\times10^{49}\,\mathrm{erg}$\footnote{Here, the incident energy of SN ejecta is obtained for a ratio of binary separation to companion radius at the moment of SN explosion of $a^{\mathrm{f}}/R^{\mathrm{f}}_{\mathrm{2}}=3.61$ (see Table~\ref{table:1}).}, is injected into the donor star during the ejecta-donor interaction. The donor star continues to expand dramatically for about 1,000 years before it starts to contract (Fig.~\ref{Fig:evolution}).  This expansion timescale is determined by the local radiative diffusion timescale of the donor envelope at the shell where it contains the injected energy from the SN Ia ejecta \citep[][see their Eq.~49]{Henyey1969}.  This expansion makes the post-impact donor star much more luminous ($\gtrsim 80\,\mathrm{L}_{\sun}$; Fig.~\ref{Fig:evolution}) than its pre-explosion brightness.  A few thousand years after the explosion, the radius and luminosity of the ejected donor star decreases again as it starts to contract, releasing gravitational energy while the deposited energy radiates away (Fig.~\ref{Fig:evolution}; Kelvin Helmholtz contraction).  Figure~\ref{Fig:track} shows the post-impact evolution tracks of the ejected donor in the Hertzsprung-Russell (H-R) diagram and in the effective temperature-surface gravity ($T_{\mathrm{eff}}-g$) diagram.

We find that including the binary orbital motion and stellar spin in 3D impact simulations have no significant effect on the post-impact evolution of the surviving donor stars (Figs.~\ref{Fig:evolution} and \ref{Fig:track}).  For a comparison, the post-impact evolution of the ejected donor has also been calculated with the \textsc{Kepler} code.  As shown in Figs.~\ref{Fig:evolution} and \ref{Fig:track}, the \textsc{Kepler} model has a lager radius thus being more luminous than the \textsc{MESA} model.  The difference between the \textsc{Kepler} and the \textsc{MESA} results is attributed to the circumstance that the output models at the end of 3D impact simulations is still somewhat out of equilibrium.  Nonetheless, the evolution trends in two approaches are very similar, and no significant difference is observed between these two models (Figs.~\ref{Fig:evolution} and \ref{Fig:track}). This indicates that at the time of the mapping from 3D to 1D the SPH models are already very close to hydrostatic equilibrium (Fig.~\ref{Fig:profile}).

\section{Discussion}
\label{sec:discussion}

\subsection{Comparison with US~708}

At the time when the SN explodes in our binary evolution calculation of Model \texttt{A} the donor star has an orbital velocity of $V_{\mathrm{orb}}=901\,\mathrm{km\,s^{-1}}$ and a surface rotational velocity of $V_{\mathrm{rot}}=320\,\mathrm{km\,s^{-1}}$ (Table~\ref{table:1}).  We directly obtain from our impact simulation that this star receives a kick velocity of about $V_{\mathrm{kick}}\approx 210\,\mathrm{km\,s^{-1}}$ due to the ejecta impact (Section~\ref{sec:impact}). This kick velocity was simply assumed to be $200\,\mathrm{km\,s^{-1}}$ by \citep{Geie15}. Therefore, we expect the donor remnant to be ejected with a velocity of $V_{\mathrm{spatial}}=(V_{\mathrm{orb}}^{2}+V_{\mathrm{kick}}^{2})^{\nicefrac12}\approx 925\,\mathrm{km\,s^{-1}}$.  This velocity is dominated by the donor's pre-explosion orbital velocity.  This spatial velocity provides a good explanation to the observed high velocity of US~708 of $\sim1,\!200\,\mathrm{km\,s^{-1}}$ \citep[Table~1 of][]{Geie15}. 

Based on the effective temperature of $T_{\mathrm{eff}}=47\mathord,200\pm 400\,\mathrm{K}$, and the surface gravity of $\mathrm{log_{10}}\,(g/\mathrm{cm\,s^{-2}})=5.69\pm 0.09$, given by \citet{Geie15}, the luminosity of US~708 can be simply calculated using $L=4\pi R^{2}{\sigma_{\mathrm{SB}}T_{\mathrm{eff}}^{4}}$ under the assumption that the mass of US~708 is $M=0.3\,\mathrm{M}_{\odot}$ \citep[see also][]{Bauer2019}.  The constant $\sigma_{\mathrm{SB}}$ is the Stefan-Boltzmann constant, as usual.  Figure~\ref{Fig:track} shows a comparison between the properties of the ejected He donor remnant, its luminosity, effective temperature, and surface gravity, as predicted by our simulations, and those observed for US~708. During its entire post-impact evolution, our Model~\texttt{A} stays too cool to pass through the positions of US~708  in the H-R diagram and in the $T_{\mathrm{eff}}-g$ diagram.  This suggests that a low-mass donor remnant ($\lesssim 0.3\,\mathrm{M}_{\odot}$) from a DDet SN Ia cannot easily  reproduce the observational properties of US~708. In the next section we explore whether a more massive ejected donor remnant from a DDet SN Ia could match the observations of US~708, including its luminosity and effective temperature.

\citet{Geie15} use the long-term evolution of a He-star without any ejecta-donor interaction to fit the observational properties of US~708.  As shown in Figs.~\ref{Fig:evolution} and~\ref{Fig:track}, the SN impact and heating during the ejecta-donor interaction play an important role in the post-impact evolution of a surviving companion star from a DDet SN Ia before it reestablishes thermal equilibrium over the Kelvin-Helmholtz timescale.  All energy deposited into the donor star by the SN blast wave has radiated away about $10^{6}\,\mathrm{yr}$ after the explosion. The star relaxes back into thermal equilibrium and continues to evolve by following an evolutionary track very close to that of a He-star with the same mass that has not been impacted and heated by the ejecta-donor interaction.  As mentioned above, based on its kinematics US~708 was ejected from the Galactic disc about $14\,\mathrm{Myr}$ ago \citep{Geie15}.  At these late epochs after the explosion, the thermal effects of ejecta-donor interaction on the evolution of the surviving He-star becomes negligible.  This suggests that the ejecta-donor interaction does not play an important role in explaining the current observational properties of US~708 if it is indeed an ejected donor remnant from a DDet SN Ia.

\subsection{More massive donor models}
\label{sec:models}

By neglecting the effect of ejecta-donor interaction, \citet{Geie15} have suggested that the evolutionary track of a post Extended Horizontal Branch (EHB) star with an original mass of $0.45\,M_{\odot}$ passes the position of US~708 in the effective temperature-surface gravity diagram. Following their suggestion, we also generate two He-star donor models with a larger initial masses to check whether the post-impact properties of a more massive ejected donor remnant in a DDet SN Ia would match the observations of US~708 better.  Model~\texttt{B} has an initial mass of $0.50\,\mathrm{M}_{\odot}$ and Model~\texttt{C} has an initial mass of $0.55\,\mathrm{M}_{\odot}$ (Table~\ref{table:1}). For both model we follow the method described in Section~\ref{sec:donor}.  The detailed binary evolution calculations are shown in Fig.~\ref{Fig:star2}.  In Model~\texttt{B} the donor star has a mass of $0.45\,\mathrm{M}_{\odot}$ at the moment of SN explosion, and Model~\texttt{C} reaches a mass of $0.50\,\mathrm{M}_{\odot}$ at SN explosion of the accreting WD.  The key properties of both models are summarised in Table~\ref{table:1}.  With these models we repeat the 3D modelling of ejecta-donor interaction and 1D post-impact evolution calculations in the same fashion as for Model~\texttt{A} and compare to the observational properties of US~708.

Figures.~\ref{Fig:evolution2} and \ref{Fig:track2} show the detailed post-impact properties and evolution tracks of Models~\texttt{B} and \texttt{C}.  These more massive surviving He-star donors reach their peak luminosities of $(1.5$--$2.1)\times10^{3}\,\mathrm{L}_{\odot}$ at $\lesssim60\,\mathrm{yr}$, much faster than the $\sim 3\mathord,000\,\mathrm{yr}$ needed in Model~\texttt{A}.  The energy deposition in the more massive models occurs at lower depths, leading to a shorter local radiative diffusion timescale of the donor envelope.  Additionally, the total amounts of energy deposition from SN ejecta in these two models are smaller than that in Model~\texttt{A} (Table~\ref{table:1}).  A fundamental and qualitative difference is that  donor models with $>0.3\,\mathrm{M}_{\odot}$ show He burning as they contract. As the deposited energy is radiated away, their temperature increases significantly (about $10^{7}\,\mathrm{yr}$ after the explosion) with an almost fixed radius and luminosity before they enter the WD cooling stage (Figs.~\ref{Fig:evolution2} and \ref{Fig:track2}).  Our most massive ejected donor remnant model ($\sim0.5\,\mathrm{M}_{\odot}$) becomes bright and hot enough to achieve the observed luminosity and temperature of US~708 \citep{Geie15}.  Unfortunately, with$\mathrm{log_{10}}\,(g/\mathrm{cm\,s^{-2}})\approx 6.20$, it has a higher surface gravity than that of US~708.  This could indicate that an even more massive He-star donor model may be needed to match the observations of US~708.  However, our detailed binary evolution calculations indicates that a binary progenitor system with such a massive He-star donor is unlikely to be close enough to reach a sufficiently high orbital velocity at the moment of SN explosion (Table~\ref{table:1}), and hence would fail to explain the observed high spatial velocity of US~708. 

Our results and conclusions are similar to those presented by \citet{Geie15}.   Figs~\ref{Fig:evolution2} and~\ref{Fig:track2} show that our surviving companion stars have re-established thermal equilibrium completely about $14\,\mathrm{Myr}$ after the explosion.  At these late epochs, the observational properties of our surviving companion stars are not sensitive to the details of ejecta-donor interaction, and are thus similar to those of EHB-star models with the same mass adopted by \citet{Geie15} at similar epochs.  This confirms that neglecting the effect of ejecta-donor interaction for the comparison with US~708 is a reasonable approximation.

Low-mass donors not becoming bright enough and high-mass donors not becoming fast enough pose problems for explaining US~708 as the He-star donor ejected from a DDet SNe Ia.  An alternative explanation  would be that the pre-explosion binary progenitor system has travelled at a speed of about $400\,\mathrm{km\,s^{-1}}$ into the direction of its current motion \citep[see also][]{Brown2015b, Bauer2019}.

\subsection{Uncertainties and future work}

In this work, the He-star donor models are constructed based on a sub-Chandrasekhar mass double-detonation explosion model given by \citet{Gronow2020}.  The exact critical He shell mass required to successfully initiate a the thermonuclear SN explosion by triggering double detonations is still quite uncertain \citep[e.g.,][]{Woosley2011,Bildsten2007,Shen2014,Townsley2019,Gronow2021}.  Our investigation is inherently limited by the assumption of a critical He-shell mass of $\sim 0.05\,\mathrm{M}_{\odot}$.  Different core and He shell masses of exploding WDs could significantly affect the SN ejecta properties and thus the predicted observational features of the resulting SNe Ia \citep[e.g.,][]{Gronow2021}.  This introduces some uncertainties into the firmness of our conclusions.   Nevertheless, we do demonstrate that certain companion masses are promising for reproducing US~708, and therefore we think that such a model seems likely to explain the main features of this star.

The exact He-retention efficiency of the accreting WD in the progenitor system is still poorly constrained \citep[e.g.,][]{Ruiter2014,Toonen2014}.  This is expected to add to the uncertainties of our 1D binary evolution calculation providing the binary properties and donor structures at the time of SN explosion.  To comprehensively model the predictions on the observables of surviving donors of DDet SNe Ia and to improve the fit with US~708, more donor models need to be constructed and investigated.  Our future work will include different DDet explosion models covering a range of core and He shell masses of exploding WDs.

In our impact simulations, we find that some SN ejecta material are captured by the donor star during the ejecta-donor interaction.  The decay of the deposited radioactive elements (e.g., $^{56}\mathrm{Ni}$) could re-heat the donor star and thus affect its post-impact evolution \citep[e.g.,][]{Shen2017}.  We leave a detailed study of the influence of captured ejecta elements on the post-impact evolution of a surviving companion star of a DDet SN Ia to a future work.

\section{Summary}
\label{sec:summary}

For the first time, we have consistently performed 3D hydrodynamical simulations of ejecta-donor interaction within the DDet scenario of SNe Ia.  We then follow the long-term post-impact evolution of surviving He-star companions of DDet SNe Ia by combining the outcomes of 3D hydrodynamic impact simulations into 1D post-impact evolution codes.  We have provided the observable signatures of the surviving  companion stars of DDet SNe Ia, which can guide the search for such companion stars with future observations. We also compare our results to the observations of the hypervelocity star US~708 to assess the validity of explaining the origin of this star with an ejected He donor from a DDet SNe Ia. The results and conclusions of this work are summarized as follows.

   \begin{enumerate}
      \item[(1)] We find that about $3\,\%$--$6\,\%$ of the initial donor mass is stripped off from their outer layers during the ejecta-donor interaction for our three He-star donor models.  The donor stars receive impact kick velocities of $80$--$200\,\mathrm{km\,s^{-1}}$, resulting in ejected donor remnants spatial velocities of $619$--$923\,\mathrm{km\,s^{-1}}$.  These velocities are dominated by the pre-explosion orbital velocities of $614$--$901\,\mathrm{km\,s^{-1}}$ (Table~\ref{table:1}).    
      
      \item[(2)] We find that the donors are significantly shocked and heated during the ejecta-donor interaction. In addition, we find that an energy of about $(1.07$--$1.15)\times10^{49}\,\mathrm{erg}$ is deposited into the donor stars, which corresponds to about $35\%$--$48\%$ of the total incident energy from SN ejecta being absorbed by the donor remnants (Table~\ref{table:1}).  The surviving donor stars inflate for about $100$--$1\mathord,000\,\mathrm{yr}$ and reach peak luminosities of $(0.08$--$2.2)\times 10^{3}\,\mathrm{L}_{\odot}$,  exceeding their pre-explosion luminosities by a up to factor of about $800$ (Figs.~\ref{Fig:evolution} and \ref{Fig:evolution2}).      
     
     \item[(3)] We find that the ejecta-donor interaction plays an important role in determining the post-impact observable signatures of surviving companion stars of DDet SNe Ia during the thermal re-equilibration phase. This leads to that the surviving companion stars become significantly overluminous for the Kelvin-Helmholtz timescale of about $10^{6}\,\rm{yr}$.  After the stars reestablish thermal equilibrium, they continue to evolve by following an evolutionary track very close to that of a He-star with the same mass that has not been impacted and heated due to the ejecta-donor interaction.
      
      \item[(4)] Although our reference model (Model~\texttt{A}) matches the high velocity of US~708 (about $1,\!200,\mathrm{km\,s^{-1}}$, \citealt{Geie15}), it does not reproduce the observed temperature and luminosity of this star. 
      
      \item[(5)]  The US~708 was ejected about $14\,\mathrm{Myr}$ ago based on its kinematics \citep{Geie15}. At these late epochs after the explosion, the observational properties of our surviving companions are not sensitive to the details of the ejecta-donor interaction. Therefore, the ejecta-donor interaction may not play an important role in examining the SN Ia origin of US~708.
      
      \item[(6)] Based on the results (Fig.~\ref{Fig:track2}) of our more massive donor star models (Model~\texttt{B} and \texttt{C}), we suggest that a He-star donor with an initial mass of $\gtrsim0.5\,\mathrm{M}_{\odot}$ is needed to reproduce the observed properties of US 708 such as its luminosity, effective temperature ans surface gravity \citep[see also][]{Geie15}. However, our detailed binary evolution calculations show that such massive donors might not reach sufficiently high orbital velocities at the moment of SN explosion in the DDet scenario to explain the observed high spatial velocity of US~708.  

      \item[(7)] Explaining the high spatial velocity of US~708 with a massive ejected donor remnant ($\gtrsim0.5\,\mathrm{M}_{\odot}$) from a DDet SN Ia event may require that its binary progenitor system has already traveled at a speed of about $400\,\mathrm{km\,s^{-1}}$ in the observed direction in the galactic halo before the SN explosion \citep[see also][]{Brown2015b}.

\end{enumerate}

To determine whether or not US~708 is indeed an ejected donor remnant from a DDet SN Ia requires further improvements of the binary evolution calculations, in particular of the accumulation efficiency of accreted He material onto the WD, and of the hydrodynamic impact modelling with different DDet explosion models covering a range of core and He shell masses of exploding white dwarfs.

\begin{acknowledgements}

We thank the anonymous referee for constructive comments that helped to improve this paper. ZWL would like to thank Robert G.~Izzard for his fruitful discussions.  ZWL is supported by the National Natural Science Foundation of China (NSFC, No.\ 11873016), the Chinese Academy of Sciences (CAS) and the Natural Science Foundation of Yunnan Province (No.\ 202001AW070007).  The work of FR is supported by the Klaus Tschira Foundation and by the Deutsche Forschungsgemeinschaft (DFG, German Research Foundation) -- Project-ID 138713538 -- SFB 881 (``The Milky Way System'', Subproject A10).  AH acknowledges support by the Australian Research Council (ARC) Centre of Excellence (CoE) for Gravitational Wave Discovery (OzGrave) project number CE170100004, by the ARC CoE for All Sky Astrophysics in 3 Dimensions (ASTRO 3D) project number CE170100013, and by the US National Science Foundation under Grant No.\ PHY-1430152 (JINA Center for the Evolution of the Elements).  The authors gratefully acknowledge the `PHOENIX Supercomputing Platform' jointly operated by the Binary Population Synthesis Group and the Stellar Astrophysics Group at Yunnan Observatories, CAS.  This work made use of the Heidelberg Supernova Model Archive (HESMA, see https://hesma.h-its.org, \citealt{Kromer2017}).

\end{acknowledgements}

\bibliographystyle{aa}

\bibliography{ref}

\end{document}